\documentclass[pra,amsmath,amssymb,aps,10pt,superscriptaddress,longbibliography,twocolumn,floatfix]{revtex4-2}

\usepackage{amsmath, amsthm, amssymb, amsfonts}
\usepackage{graphicx} 
\usepackage[colorlinks=true, linkcolor=blue, urlcolor=blue, citecolor=blue, anchorcolor = blue]{hyperref} 
\usepackage{color,xcolor}

\usepackage{dcolumn}
\newcolumntype{d}[1]{D{.}{.}{#1}}

\newlength{\lengthofminus}
\newcommand{\dl}{\hspace{\lengthofminus}} 

\newlength{\lengthofone}
\newcommand{\dd}{\hspace{\lengthofone}} 
\begin{document}
\title{Microwave transitions in atomic sodium: \\ Radiometry and polarimetry using the sodium layer}

%\date{\today}

\author{Mariusz Pawlak}
\email{e-mail: teomar@chem.umk.pl}
\affiliation{ 
Faculty of Chemistry, Nicolaus Copernicus University in Toru\'n, Gagarina~7, 87-100~Toru\'n, Poland
}
\author{Eve L. Schoen}
\affiliation{
Department of Physics, University of California, 366 Physics North MC 7300, Berkeley, California 94720, USA
}
\affiliation{
MIT Kavli Institute for Astrophysics and Space Research, Massachusetts Institute of Technology, 77 Massachusetts Ave, Cambridge, Massachusetts 02139, USA
}
\author{Justin E. Albert}
\affiliation{Department of Physics and Astronomy, University of Victoria, Victoria, British Columbia V8P 5C2, Canada 
}
\author{H. R. Sadeghpour}
\affiliation{
ITAMP, Center for Astrophysics $|$ Harvard $\&$ Smithsonian,  60 Garden St, Cambridge, Massachusetts 02138, USA
}

%%%%%%%%%%%%%%%%%%%%%%%%%%%%%%%%%%%%%%%%%%%%%%%%%%%%%%%%%%%%%%%%%%%%%%%%%%%%
\begin{abstract}
We calculate, via variational techniques, single- and two-photon Rydberg microwave transitions, as well as scalar and tensor polarizabilities of sodium atom using the parametric one-electron valence potential, including the spin--orbit coupling. The trial function is expanded in a~basis set of optimized Slater-type orbitals, resulting in highly accurate and converged eigen-energies up to $n=60$. We focus our studies on the microwave band 90--150~GHz, due to its relevance to laser excitation in the Earth's upper-atmospheric sodium layer for wavelength-dependent radiometry and polarimetry, as precise microwave polarimetry in this band is an important source of systematic uncertainty in searches for signatures of primordial gravitational waves within the anisotropic polarization pattern of photons from the cosmic microwave background. We present the most efficient transition coefficients in this range, as well as the scalar and tensor polarizabilities compared with available experimental and theoretical data.  
\end{abstract}
\maketitle

%%%%%%%%%%%%%%%%%%%%%%%%%%%%%%%%%%%%%%%%%%%%%%%%%%%%%%%%%%%%%%%%%%%%%%%%%%%%%
\section{Introduction}
The frequency domain in the MHz--THz range is a~preferred band for sensing and metrology in the radio to millimeter wavelengths, as well as for establishing measurement standards. One of the most promising and sensitive recent developments in this field is with the use of Rydberg atom electric field sensing~\cite{Osterwalder_Merkt_1999, Sedlacek2012, Sassmannshausen_PRA_2013, Fan_OptLett_2014, Holloway_APL_2014, Gordon_APL_2014, Fan_JPB_2015, Fan_PRApp_2015, Simons_APL_2016, Degen_2017, Holloway_IEEE_2017, Sapiro_2020, Raithel2022, Yuan_2023}. This appealing aspect allows for amplitude sensing of RF fields~\cite{Sapiro_2020} and for imaging with sub-wavelength spatial resolution~\cite{Fan_OptLett_2014, Holloway_IEEE_2017}.

A~major source of uncertainty in searches for signatures of primordial gravitational waves within the polarization pattern of the cosmic microwave background (CMB), and a~source that is growing in relative importance as the precision of such searches improves, is the lack of microwave sources of precisely-known polarization in the sky to use for calibration~\cite{Stubbs2006}. Recently, excitations with lasers in the Earth's sodium layer were shown to dramatically improve photometric systematics, via a~laser photometric ratio star (LPRS) in the visible and near-infrared frequencies~\cite{Albert2021a,Albert2021b,Albert2022}. It may be possible to extend the LPRS to precise calibration of microwave telescopes, such as the BICEP/Keck Array, which measure the polarization anisotropy of the CMB, and future enlarged and improved microwave observatories such as CMB-S4~\cite{Albert2022b,CMB_S4}. Polarized microwave and millimeter-wave radiation from Rydberg excited atoms in the sodium layer could be used for precise relative radiometric and polarimetric calibrations of such ground-based microwave and millimeter-wave telescopes.

\begin{figure}[b]
\centering
\includegraphics[width=8.1cm]{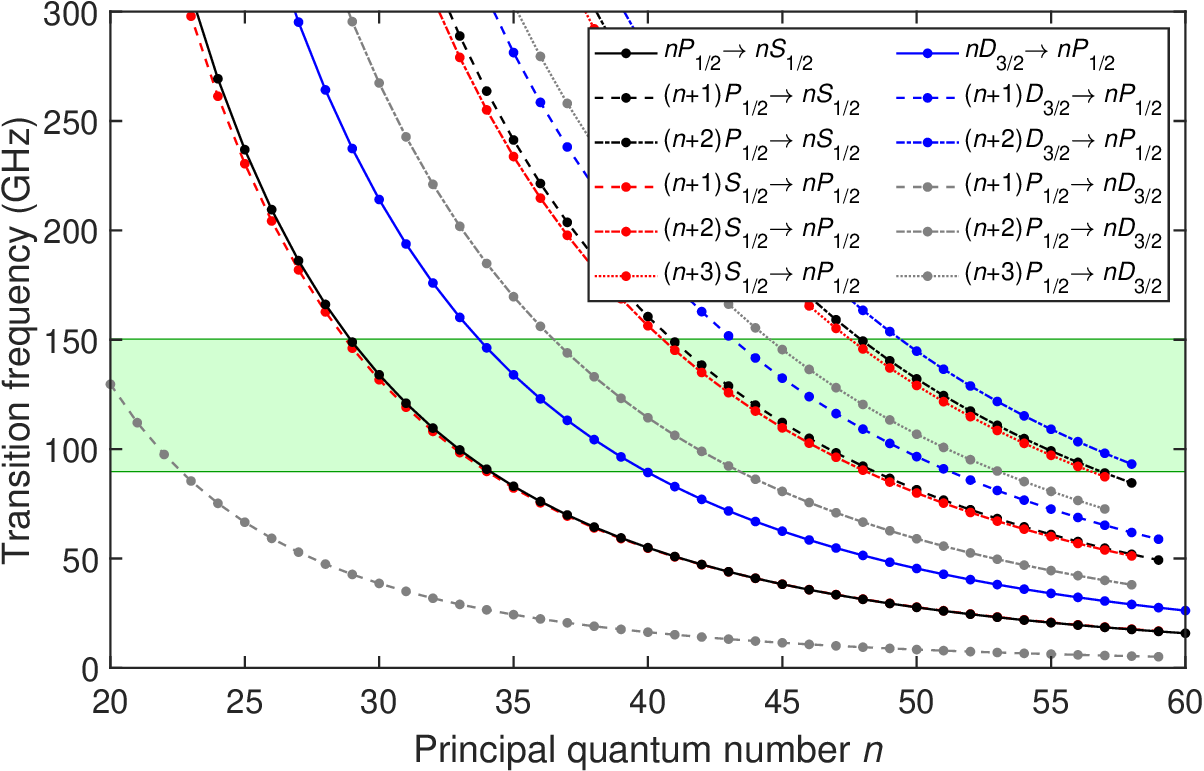}
\caption{\label{fig:band} 
The $nP_{1/2}\leftrightarrow n^\prime S_{1/2}$ and $nD_{3/2}\leftrightarrow n^\prime P_{1/2}$ electron transitions in the highly excited Na atom. The green area indicates the microwave frequency band of 90 to 150~GHz. The spectrum for the $nP_{3/2}\leftrightarrow n^\prime S_{1/2}$ and $nD_{3/2,5/2}\leftrightarrow n^\prime P_{3/2}$ electron transitions looks essentially the same.}
\end{figure}

Precise measurements and calculations of atomic polarizability play an essential role not only in metrology but for quantum information processing, optical trapping and cooling, and interparticle collision studies. The Einstein $A$~coefficients are especially required for applications in atmospheric physics and astrophysics. There is a~great need to know these quantities with high accuracy over a~wide range of frequencies.

The primary source of recommended data for atomic parameters of the sodium atom is the report of NIST~\cite{Kelleher,NIST_ASD}. Kelleher and Podobedova \cite{Kelleher} complied frequencies and spontaneous emission coefficients (Einstein $A$~coefficients) for transitions in Na up to $n=11$ and angular momentum $0\le l \le 3$. It is worth noting that Gallagher {\it et al.}~\cite{Gallagher_1977} made radio-frequency resonance measurements of the $nP$ and $nD$ series for $n=16$--19 and $n=15$--17, respectively. Fabre, Haroche, and Goy~\cite{Fabre_Haroche_1978} observed more highly excited states of Na and measured the microwave resonance frequencies between $S$, $P$, $D$, and $F$ states for $n=23$--41 as well as the polarizabilities of $nS$ and $nP$ levels.

In this article, we extend the previous data on Einstein coefficients and scalar and tensor polarizabilities of Rydberg states of the sodium atom to include higher principal quantum numbers. Based on an accurate variational approach~\cite{Pawlak_PRA_2014,Pawlak_PRA_2020}, transition frequencies, spontaneous and stimulated emission coefficients, and polarizabilities for the Na($n\le60$, $l \le 3$) atom are calculated. We particularly focus on the prominent lines in the 90--150~GHz range, of interest for precision microwave relative radiometric and polarimetric laser excitation of the sodium layer~\cite{Albert2021a,Albert2021b,Albert2022}. Rydberg transitions within Earth's mesosphere can offer precise sources of relative radiometry and polarimetry in the microwave, with the use of free-space lasers~\cite{dang22}. The green area in Fig.~\ref{fig:band} shows the microwave band under consideration within the calculated transition spectrum of Na. One can see the abundance of atomic electron transitions we will concentrate on in this article.

%%%%%%%%%%%%%%%%%%%%%%%%%%%%%%%%%%%%%%%%%%%%%%%%%%%%%%%%%%%%%%%%%%%%%%%%%%%%%
\section{Theory and computation}

\subsection{Hamiltonian terms}
We consider a~system comprised of the valence electron interacting with a~closed alkali-metal positive-ion core. The Hamilton operator can be written as (in a.u.)
\begin{equation} \label{Hamil}
\hat H = -\frac{1}{2}\nabla^2 + V_l(r) + V_{\rm LS}(r),     
\end{equation}
where $V_l(r)$ is the one-dimensional effective model potential~\cite{Marinescu_1994} and $V_{\rm LS}(r)$ is the spin--orbit coupling~\cite{Aymar_1996}. The form of $V_l(r)$ depends on the nuclear charge ($Z$) of the atom, five parameters ($a_k^{(l)}$, $k=1, 2, 3, 4$, and $r_c^{(l)}$), and the static dipole polarizability ($\alpha_c$) of the singly charged core, 
\begin{eqnarray}
 V_l(r) & = & -\frac{1}{r} -\frac{(Z-1)}{r} \exp \left [-a_1^{(l)} r \right ] \nonumber \\ 
 & & + \left  (a_3^{(l)}+a_4^{(l)} r \right) \exp \left [-a_2^{(l)} r \right ]  \nonumber \\
 & & - \frac{\alpha_c}{2r^4} \left ( 1-\exp \left [-\left (r/r_c^{(l)} \right )^6 \right ] \right ). 
\end{eqnarray}
All the parameters for the sodium atom are given in Ref.~\cite{Marinescu_1994}. The spin--orbit interaction potential may be expressed as (in a.u.)
\begin{equation} \label{potVLS}
V_{\rm LS}(r)=  {\textbf{L}}\cdot{\textbf{S}} \frac{\alpha^2}{2} \frac{1}{r}\frac{{\rm d}V_l(r)}{{\rm d}r}  \left (1 -\frac{\alpha^2}{2}V_l(r)  \right )^{-2}, 
\end{equation}
where $\langle {\textbf{L}}\cdot{\textbf{S}} \rangle = \frac{1}{2}[j(j+1)-l(l+1)-\frac{3}{4}]$ and $\alpha$ stands for the fine-structure constant. As usual, $j$ and $l$ are the total and orbital angular momentum quantum numbers, respectively, where $j=l\pm \frac{1}{2}.$

\begin{figure}[t] 
\centering
\includegraphics[width=8.1cm]{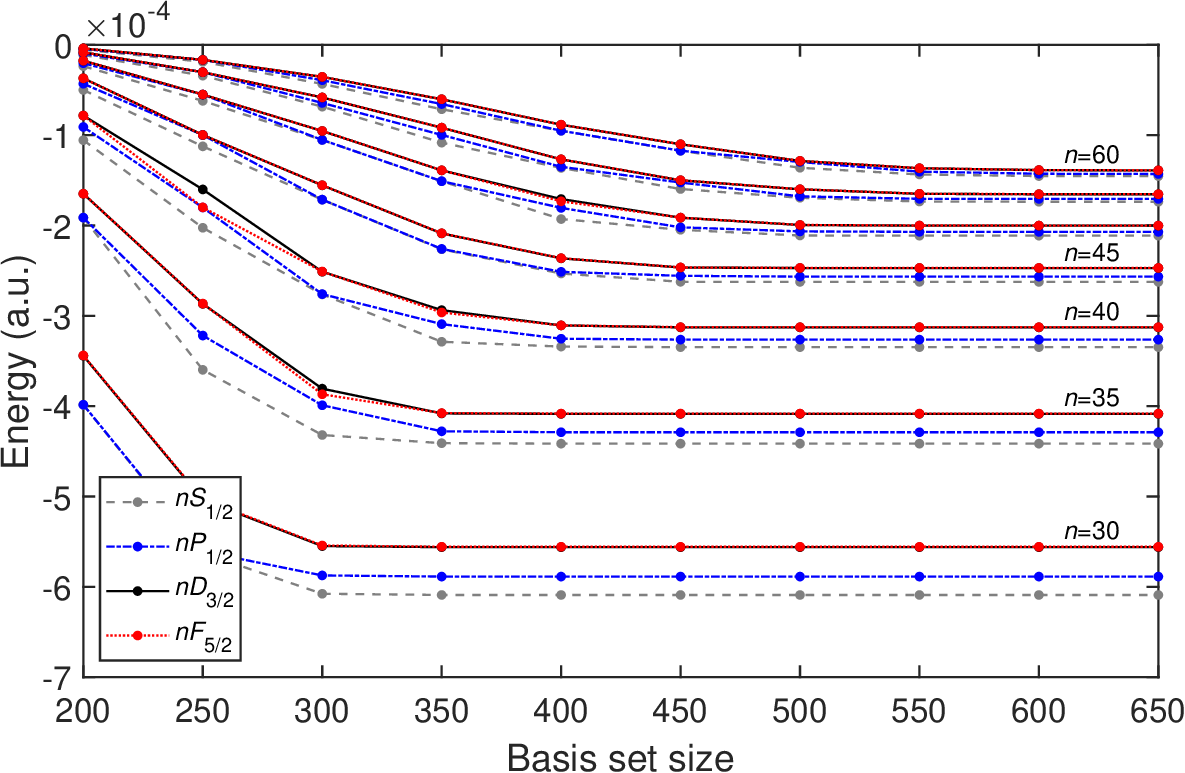}
\caption{ \label{fig:basis_set}
Convergence of $nS_{1/2}$, $nP_{1/2}$, $nD_{3/2}$, and $nF_{5/2}$ energy levels with basis-set size, for different $n$.}
\end{figure}

We compute the eigen-spectra of the time-independent Schr\"{o}dinger equation with the Hamiltonian~(\ref{Hamil}) for Na($l\le 3$, $j$) by applying the Ritz variational method in a~trial space spanned by 650 optimized Slater-type orbitals (STOs). Technical details of the calculations are described in sections II of Ref.~\cite{Pawlak_PRA_2014} and II.B of Ref.~\cite{Pawlak_PRA_2020}. The calculations are carried out in quadruple precision. According to the variational principle for excited states, the lowest $k$-th eigen-energy $(E_k)$ is the upper bound for the respective exact energy of the $k$-th state of Na. Any increase in the size of the basis set gives an improved approximation (i.e., lower energy) $E_k$ for the $k$-th exact energy level. The convergence of selected energies with respect to the size of the basis set is displayed in Fig.~\ref{fig:basis_set}. Saturation of the obtained results is visible when the trial space is enlarged. However, the higher the energy level, the slower convergence to the exact solution. The variational method with 650 STOs enables us to compute the energy spectrum for any $(l,j)$ series up to $n = 60$ with accuracy not less than four significant digits (for the worst case), and the accuracy increases with decreasing~$n$.

\begin{table*}[th!] 
\caption {\label{tab:polariz} The convergence of the scalar and tensor polarizabilities (in a.u.), for the lowest states and some of the sodium Rydberg states with respect to $n^\prime$, in the sums of Eqs.~(\ref{eqn:scalar}) and~(\ref{eqn:tensor}). The notation $(x)$ denotes $\times 10^x$. The closed interval $[a,b]$ is defined by $\{n^\prime\in \mathbb{N}: a \le n^\prime \le b \} $.
}
\centering
\begin{tabular}{cc@{\qquad}llllllll}
\hline\hline
$n$ &  $n^\prime$   & \multicolumn{1}{c}{$\alpha_0(S_{1/2})$} & 
                      \multicolumn{1}{c}{$\alpha_0(P_{3/2})$} & 
                      \multicolumn{1}{c}{$\alpha_0(D_{5/2})$} & 
                      \multicolumn{1}{c}{$\alpha_2(P_{3/2})$} & 
                      \multicolumn{1}{c}{$\alpha_2(D_{5/2})$} \\
\hline
3  & $\le 60$  & 1.66919(2)  & \dl 3.56032(2) & 6.38605(3)  & $-8.45922(1)$   & $-5.07481(3)$ \\
   & $\le 50$  & 1.66919(2)  & \dl 3.56023(2) & 6.38596(3)  & $-8.45900(1)$   & $-5.07478(3)$ \\
   & $\le 40$  & 1.66919(2)  & \dl 3.56005(2) & 6.38580(3)  & $-8.45859(1)$   & $-5.07474(3)$ \\
   & $\le 30$  & 1.66919(2)  & \dl 3.55966(2) & 6.38544(3)  & $-8.45768(1)$   & $-5.07463(3)$ \\
   & $\le 20$  & 1.66918(2)  & \dl 3.55853(2) & 6.38439(3)  & $-8.45501(1)$   & $-5.07434(3)$ \\
   & $\le 10$  & 1.66914(2)  & \dl 3.55191(2) & 6.37784(3)  & $-8.43823(1)$   & $-5.07246(3)$ \\
\\ 
30 & $\le 60$  & 6.78051(9)  & $-9.70554(10)$ & 1.62196(12) & \dl 9.81951(9)  & $-5.16578(11)$ \\
   & $\le 50$  & 6.78048(9)  & $-9.70555(10)$ & 1.62196(12) & \dl 9.81954(9)  & $-5.16578(11)$ \\
   & $\le 40$  & 6.78031(9)  & $-9.70562(10)$ & 1.62196(12) & \dl 9.81972(9)  & $-5.16578(11)$ \\
   & $\le 35$  & 6.77951(9)  & $-9.70593(10)$ & 1.62196(12) & \dl 9.82066(9)  & $-5.16576(11)$ \\
\\
   & $[10,60]$ & 6.78051(9)  & $-9.70554(10)$ & 1.62196(12) & \dl 9.81951(9)  & $-5.16578(11)$ \\
   & $[20,60]$ & 6.78052(9)  & $-9.70554(10)$ & 1.62196(12) & \dl 9.81951(9)  & $-5.16578(11)$ \\
   & $[25,60]$ & 6.78091(9)  & $-9.70552(10)$ & 1.62196(12) & \dl 9.81945(9)  & $-5.16578(11)$ \\
\\
50 & $\le 60$  & 1.94641(11) & $-3.85495(12)$ & 5.82569(13) & \dl 3.98254(11) & $-1.85504(13)$ \\
   & $\le 55$  & 1.94615(11) & $-3.85504(12)$ & 5.82568(13) & \dl 3.98282(11) & $-1.85503(13)$ \\
\\
   & $[10,60]$ & 1.94641(11) & $-3.85495(12)$ & 5.82569(13) & \dl 3.98254(11) & $-1.85504(13)$ \\
   & $[20,60]$ & 1.94641(11) & $-3.85495(12)$ & 5.82569(13) & \dl 3.98254(11) & $-1.85504(13)$ \\
   & $[30,60]$ & 1.94641(11) & $-3.85495(12)$ & 5.82569(13) & \dl 3.98254(11) & $-1.85504(13)$ \\
   & $[40,60]$ & 1.94642(11) & $-3.85495(12)$ & 5.82569(13) & \dl 3.98254(11) & $-1.85504(13)$ \\
   & $[45,60]$ & 1.94665(11) & $-3.85494(12)$ & 5.82570(13) & \dl 3.98250(11) & $-1.85504(13)$ \\
\hline\hline
\end{tabular}
\end{table*}

%%%%%%%%%%%%%%%%%%%%%%%%%%%%%%%%%%%%%%%%%%%%%%%%%%%%%%%%%%%%%%%%%%%%%%%%%%%%%
\subsection{Polarizabilities}
The polarizability of an atom is a measure of the atomic orbital distortion, exposed to an applied electric field. In the case of linearly polarized electric field~$F$, the interaction with an atom in a~state described by the quantum numbers $j$ (total angular momentum) and $m_j$ (the projection of $j$ on the axis of propagation of the electromagnetic wave) is written as~\cite{vanWij_1994}
\begin{equation}
V_F = -\frac{1}{2} \left ( \alpha_0 + \alpha_2\frac{3m_j^2-j(j+1)}{j(2j-1)}  \right ) F^2.
\end{equation}
We calculate the scalar and tensor polarizabilities, $a_0$ and $a_2$, respectively, of Na$(n,l,j)$ states as follows~\cite{Khadjavi_1968, Lai_polarizabilities}
\begin{eqnarray} 
  a_0 & = &  -\frac{2}{3} \sum_{n',l',j'} (2j'+1)   
     \begin{Bmatrix} 
        l & j & 1/2  \\
        j' & l' & 1  \\
     \end{Bmatrix}^2  \nonumber \\
    & & \times \text{max}(l, l') \frac{ | \langle nl |r | n'l' \rangle |^2}{E_{n,l,j}-E_{n',l',j'}},  \label{eqn:scalar} \\
  a_2 & = & -2 \left[ \frac{10j(2j-1)(2j+1)}{3(j+1)(2j+3)} \right]^{1/2}\sum_{n',l',j'} (-1)^{j+j'}(2j'+1) \nonumber \\
 & &
   \times     \text{max}(l, l') 
    \begin{Bmatrix} 
   l & j & 1/2  \\
   j' & l' & 1  \\
   \end{Bmatrix}^2
   \begin{Bmatrix} 
   j & j' & 1  \\
   1 & 2 & j  \\
   \end{Bmatrix}
   \frac{ | \langle nl |r | n'l' \rangle |^2}{E_{n,l,j}-E_{n',l',j'}}. \label{eqn:tensor}  \nonumber  \\
\end{eqnarray}
The curly brackets, $\{\,:\,:\,:\,\}$, refer to a~Wigner 6$j$ symbol. In theory, the summations should be carried out over all bound and continuum states, but in practice, only the neighboring states contribute significantly~\cite{Lai_polarizabilities}. Obviously, the dipole-allowed transitions are for $l^\prime = l \pm 1$. Since we successfully calculated the Na($nS_{1/2}$, $nP_{1/2,3/2}$, $nD_{3/2,5/2}$, $nF_{5/2,7/2}$) energy levels up to $n=60$, we were able to determine the polarizabilities for $nS_{1/2}$, $nP_{1/2,3/2}$, and $nD_{3/2,5/2}$ series with $n\le58$ with high accuracy. Note that the radial dipole matrix elements, $ \langle nl |r| n'l' \rangle$, do not depend on $j$, i.e., the wave functions describe the $(n,l)$ states, where spin--orbit coupling is not incorporated. In turn, energy levels in the above equations include the fine-structure splitting due to the spin--orbit interaction.

The quality of the polarizability computations in the range we are interested in is quite satisfactory. Table~\ref{tab:polariz} shows convergence of polarizabilities with respect to the number of terms in evaluating Eqs.~(\ref{eqn:scalar}) and (\ref{eqn:tensor}) for selected states.  It is plainly seen that the polarizabilities are contributed by the most adjacent intermediate states. Let us take a~closer look at the results for $\alpha_0(P_{3/2})$ as an example. The scalar polarizability for $3P_{3/2}$ is equal to 355.191~a.u.\ when the summation in Eq.~(\ref{eqn:scalar}) is over neighboring states $(n^\prime\le 10)$, and 356.032~a.u.\ when more states are taken into account $(n^\prime\le 60)$. The relative error is less than 0.24\%. Moreover, one can see in Table~\ref{tab:polariz} how the results saturate when the upper limit of the summation systematically increases (from 10 to 60 with step 10). The convergence of the scalar polarizability for excited states is even better. The relative error of $\alpha_0(50P_{3/2})$ when $45\le n^\prime \le 60$ compared to the case with $n^\prime\le 60$ is less than 0.0003\%. Similar observations can be made for other polarizabilities presented in Table~\ref{tab:polariz}.

%%%%%%%%%%%%%%%%%%%%%%%%%%%%%%%%%%%%%%%%%%%%%%%%%%%%%%%%%%%%%%%%%%%%%%%%%%%%%
\subsection{Einstein coefficients} 
The Einstein coefficients for atom--photon reactions are intrinsic properties of any atom that do not depend on the nature of the external electromagnetic radiation.

The Einstein $A$ coefficient corresponds to the rate of spontaneous emission from a~higher energy state ($k$) to a~lower energy state ($i$), $A_{k,i} \equiv  A[(n,l,j) \rightarrow (n',l',j')]$. Using this notation, we have (in a.u.)~\cite{Miculis_2005} 
\begin{equation}\label{eqn:A}
    A_{k,i} = \frac{4}{3(2j+1)}\alpha^3  E_{k,i}^3 | \langle nlj||D(r)||n'l'j'\rangle|^2,
\end{equation}
where $E_{k,i}$ is the energy of the transition, $E_{k,i} = E_k - E_i = E_{n,l,j} - E_{n^\prime, l^\prime, j^\prime}  = \hbar \omega_{k,i}$. The square of the matrix element of the transition dipole operator, $D(r)$, is given by \cite{Miculis_2005,Sobelman_book}
\begin{eqnarray}\label{eqn:dipole}
  | \langle n l j || D(r) || n 'l' j' \rangle |^2 & = & (2j+1)(2j'+1) \begin{Bmatrix}
 l && j && 1/2 \\
 j' && l' && 1 \\
\end{Bmatrix}^2  \nonumber \\ 
&& \times \text{max}(l, l')|\langle nl |r | n'l' \rangle|^2.   
\end{eqnarray}

The Einstein $B$ coefficient describes the rate of stimulated emission and absorption per time and energy density of the incident radiation. Stimulated emission can be related to spontaneous emission by the inverse cube of the transition frequency from the initial (upper) to the final (lower) state by (in~a.u.)~\cite{Woodgate_book, Dodd_1991}
\begin{equation} \label{eqn:B1}
    B_{k,i} = \frac{c^3 \pi^2}{ \omega_{k,i}^3 }A_{k,i} .
\end{equation}
Incorporating Eqs.~(\ref{eqn:A}) and (\ref{eqn:dipole}) into Eq.~(\ref{eqn:B1}), we get a~final equation for $B_{k,i}$, which does not depend directly on the energy level difference (in~a.u.)
\begin{eqnarray}
B_{k,i} &=& \frac{4}{3}\pi^2  (2j'+1) 
    \begin{Bmatrix}
     l && j && 1/2 \\
     j' && l' && 1 \\
    \end{Bmatrix}^2  \nonumber \\
& & \times \text{max}(l, l') |\langle nl |r | n'l' \rangle|^2.
\end{eqnarray}
The stimulated absorption rate coefficient, $B_{i,k}$, is proportional to $B_{k,i}$ by the ratio of the statistical weights~\cite{Woodgate_book, Dodd_1991},
\begin{equation} \label{BB}
     B_{i,k} = \frac{g_k}{g_i} B_{k,i}.
\end{equation}

%%%%%%%%%%%%%%%%%%%%%%%%%%%%%%%%%%%%%%%%%%%%%%%%%%%%%%%%%%%%%%%%%%%%%%%%%%%%%
\subsection{Two-photon transitions}
The probability per second to emit two photons, one at frequency $\omega_1$ and the other at $\omega_2$, in the transition $nS_{1/2}\rightarrow n'P_{j'} \rightarrow  n^{\prime\prime} S_{1/2} $, $j^\prime=\frac{1}{2},\frac{3}{2}$, is (in a.u.)~\cite{Breit_1940, Demidov_1962, Dalgarno_1966} 
\begin{equation}\label{lab_Ay}
        A(y)dy = \frac{8}{27\pi} \alpha^6 \omega_{n,n''}^7 y^3 (1-y)^3   \left | M_{j^\prime} \right |^2 dy,
\end{equation}
where $y = \omega_1/\omega_{n,n''}$, $\omega_{n,n''}=\omega_1+\omega_2$, and
\begin{eqnarray}
  M_{j^\prime} &=& M_{j^\prime}^{(1)}+M_{j^\prime}^{(2)}  = \sum_{n'} \langle nS |r | n'P \rangle \langle n'P |r | n''S \rangle  \nonumber \\
  && \times \left( \frac{1}{E_{n'P_{j'}}-E_{nS_{1/2}} + y (E_{nS_{1/2}}-E_{n''S_{1/2}})} \right .  \nonumber \\ 
    && + \left . \frac{1}{E_{n'P_{j'}}-E_{nS_{1/2}} + (1-y) (E_{nS_{1/2}}-E_{n''S_{1/2}})} \right )  \nonumber . \\
    \label{lab_Mj}
\end{eqnarray}
In the above, $\hbar\omega_{n,n''}$ is the transition energy from an upper state $(n,l,j)$ to a lower state $(n'',l'',j'')$, $E_{n,l,j}-E_{n'',l'',j''}$, and the sum is performed over all intermediate dipole-allowed atomic states $n^\prime$.

The total probability for the two-photon emission is 
\begin{equation}
    A_t=\frac{1}{2} \int_0^1 A(y)dy.
\end{equation}
The factor in front of the integral avoids double counting of photons. Consequently, the mean life time is $\tau=A_t^{-1}$.

%%%%%%%%%%%%%%%%%%%%%%%%%%%%%%%%%%%%%%%%%%%%%%%%%%%%%%%%%%%%%%%%%%%%%%%%%%%%%

\begin{table*}[p]
\caption {\label{tab_polar} Selected scalar (given by $\alpha_0$) and tensor (given by $\alpha_2$) polarizabilities of the $nS$, $nP$, and $nD$ 
states of Na in a.u. The notation $(x)$ denotes $\times 10^x$.}
\begin{tabular}{c@{\qquad\quad}l@{\quad}ll@{\quad}ll@{\qquad}l@{\quad}ll}
\hline \hline
$n$ & \multicolumn{1}{l}{$\alpha_0(S_{1/2})$} & \multicolumn{1}{l}{$\; \alpha_0(P_{1/2})$} & \multicolumn{1}{l}{$\;\; \alpha_0(P_{3/2})$} 
    & \multicolumn{1}{l}{$\!\! \alpha_0(D_{3/2})$} & \multicolumn{1}{l}{$\alpha_0(D_{5/2})$} & \multicolumn{1}{l}{$\alpha_2(P_{3/2})$} 
    & \multicolumn{1}{l}{$\alpha_2(D_{3/2})$} & \multicolumn{1}{l}{$\alpha_2(D_{5/2})$}\\
\hline                                                      
3 & $1.669(2)$   &  $\dl 3.556(2)  $ & $\dl 3.560(2)  $ & $6.399(3)  $ & $6.386(3)  $         & $   -8.459(1)  $ & $-3.565(3)  $ & $-5.075(3)  $  \\
  & $1.655(2)^a$ &  \multicolumn{2}{c}{$3.607(2)^b$}    &\multicolumn{2}{c}{$6.396(3)^b\quad$}& $   -8.789(1)^b$ &               & $-5.073(3)^b$  \\
  & $1.673(2)^c$ &  $\dl 3.449(2)^c$ & $\dl 3.462(2)^c$ & $6.646(3)^c$ & $6.622(3)^c$         & $   -9.966(1)^c$ & $-3.723(3)^c$ & $-5.285(3)^c$  \\
  & $1.627(2)^d$ &  $\dl 3.597(2)^e$ & $\dl 3.614(2)^e$ &              &                      & $   -8.80(1)^e $ &               &                \\
  & $1.59(2)^f $ &  $\dl 3.49(2)^g$  &                  &              &                      & $   -1.13(2)^g $ &               &                \\
4 & $3.132(3)$   &  $   -4.517(3)  $ & $   -4.486(3)  $ & $6.112(5)  $ & $6.113(5)  $         & $   -1.361(2)  $ & $-1.443(5)  $ & $-2.060(5)  $  \\
  & $3.309(3)^c$ &  $   -4.694(3)^c$ & $   -4.628(3)^c$ & $6.365(5)^c$ & $6.358(5)^c$         & $   -3.565(2)^c$ & $-1.501(5)^c$ & $-2.140(5)^c$  \\
  & $3.110(3)^b$ &                   &                  & $6.109(5)^h$ & $6.107(5)^h$         &                  & $-1.432(5)^h$ & $-2.052(5)^h$  \\
  &              &                   &                  & $6.241(5)^i$ & $6.273(5)^i$         &                  & $-1.547(5)^i$ & $-2.138(5)^i$  \\
5 & $2.191(4)$   &  $   -5.867(4)  $ & $   -5.839(4)  $ & $4.009(6)  $ & $4.010(6)  $         & $\dl 2.674(3)  $ & $-9.174(5)  $ & $-1.310(6)  $  \\
  & $2.352(4)^c$ &  $   -5.990(4)^c$ & $   -5.933(4)^c$ & $4.065(6)^c$ & $4.060(6)^c$         & $\dl 1.387(3)^c$ & $-9.317(5)^c$ & $-1.328(6)^c$  \\
  & $2.33(4)^h$  &                                                                                                                                \\
  & $2.1(4)^i$   &                                                                                                                                \\
10& $4.137(6)  $ & $    -2.476(7)  $ & $   -2.467(7)  $ & $6.900(8)  $ & $6.902(8)  $         & $\dl 2.167(6)  $ & $-1.545(8)  $ & $-2.207(8)  $  \\
  &              &                   &                  &              &                      &                  &               & $-2.1(8)^j  $  \\
  &              &                   &                  &              &                      &                  &               & $-2.8(8)^k  $  \\
11& $8.069(6)  $ & $    -5.262(7)  $ & $   -5.243(7)  $	& $1.364(9)  $ & $1.365(9)  $        & $\dl 4.719(6)  $ & $-3.053(8)  $ & $-4.360(8)  $  \\
  &              &                   &                  &              &                      &                  &               & $-4.22(8)^j $  \\
  &              &                   &                  &              &                      &                  &               & $-4.82(8)^k $  \\
12& $1.475(7)  $ & $    -1.038(8)  $ & $   -1.034(8)  $	& $2.536(9)  $ & $2.537(9)  $        & $\dl 9.486(6)  $ & $-5.671(8)  $ & $-8.099(8)  $  \\
  &              &                   &                  &              &                      &                  &               & $-7.84(8)^j $  \\
  &              &                   &                  &              &                      &                  &               & $-8.24(8)^k $  \\
15& $6.767(7)  $ &  $   -5.749(8)  $ & $   -5.728(8)  $ & $1.234(10) $ & $1.234(10) $         & $\dl 5.451(7)  $ & $-2.756(9)  $ & $-3.936(9)  $  \\
  &              &                   &                  &\multicolumn{2}{c}{$1.19(10)^l\quad$}&                  &               & $-3.8(9)^l  $  \\
  &              &                   &                  &              &                      &                  &               & $-4.26(9)^m $  \\
16& $1.046(8)  $ &  $   -9.367(8)  $ & $   -9.333(8)  $ & $1.947(10) $ & $1.948(10) $         & $\dl 8.958(7)  $ & $-4.348(9)  $ & $-6.210(9)  $  \\
  &              &  \multicolumn{2}{c}{$-9.00(8)^l$}    &\multicolumn{2}{c}{$1.89(10)^l\quad$}& $\dl 7.96(7)^l $ &               & $-5.99(9)^l $  \\
  &              &                   &                  &              &                      & $\dl 8.8(7)^m  $ &               & $-5.91(9)^m $  \\
17& $1.572(8)  $ &  $   -1.478(9)  $ & $   -1.473(9)  $ & $2.987(10) $ & $2.988(10) $         & $\dl 1.424(8)  $ & $-6.669(9)  $ & $-9.524(9)  $  \\
  &              &  \multicolumn{2}{c}{$-1.40(9)^l$}    &\multicolumn{2}{c}{$2.88(10)^l\quad$}& $\dl 1.26(8)^l $ &               & $-9.20(9)^l $  \\
  &              &                   &                  &              &                      & $\dl 1.3(8)^m  $ &               & $-1.12(10)^m$  \\
18& $2.305(8)  $ &  $   -2.268(9)  $ & $   -2.260(9)  $ & $4.470(10) $ & $4.471(10) $         & $\dl 2.198(8)  $ & $-9.978(9)  $ & $-1.425(10) $  \\
  &              &  \multicolumn{2}{c}{$-2.18(9)^l$}    &              &                      & $\dl 1.97(8)^l $ &               &                \\
  &              &                   &                  &              &                      & $\dl 2.0(8)^m  $ &               &                \\
19& $3.307(8)  $ &  $   -3.394(9)  $ & $   -3.382(9)  $ & $6.543(10) $ & $6.544(10) $         & $\dl 3.309(8)  $ & $-1.460(10) $ & $-2.085(10) $  \\
  &              &  \multicolumn{2}{c}{$-3.25(9)^l$}    &              &                      & $\dl 2.98(8)^l $ &               &                \\
  &              &                   &                  &              &                      & $\dl 3.0(8)^m  $ &               &                \\
20& $4.653(8)  $ &  $   -4.969(9)  $ & $   -4.952(9)  $ & $9.389(10) $ & $9.391(10) $         & $\dl 4.868(8)  $ & $-2.095(10) $ & $-2.992(10) $  \\
23& $1.176(9)  $ &  $   -1.396(10) $ & $   -1.391(10) $ & $2.509(11) $ & $2.510(11) $         & $\dl 1.383(9)  $ & $-5.599(10) $ & $-7.996(10) $  \\
  & $(\ast)    $ & \multicolumn{2}{c}{$-1.4(10)^n$}     &              &                      & $\dl 1.4(9)^n  $ &               &                \\
  & $(\ast\ast)$ & \multicolumn{2}{c}{$-1.7(10)^o$}     &              &                      & $\dl 1.8(9)^o  $ &               &                \\
24& $1.558(9)  $ &  $   -1.908(10) $ & $   -1.902(10) $ & $3.384(11) $ & $3.385(11) $         & $\dl 1.897(9)  $ & $-7.551(10) $ & $-1.078(11) $  \\
25& $2.040(9)  $ & $    -2.575(10) $ & $   -2.566(10) $ & $4.509(11) $ & $4.509(11) $         & $\dl 2.567(9)  $   & $-1.006(11) $ & $-1.436(11) $\\
30& $6.781(9)  $ & $    -9.740(10) $ & $   -9.706(10) $ & $1.622(12) $ & $1.622(12) $         & $\dl 9.820(9)  $ & $-3.617(11) $ & $-5.166(11) $  \\
32& $1.036(10) $ & $    -1.556(11) $ & $   -1.551(11) $ & $2.550(12) $ & $2.551(12) $         & $\dl 1.574(10) $ & $-5.689(11) $ & $-8.124(11) $  \\
  &              & \multicolumn{2}{c}{$-1.6(11)^n \,\;$}&              &                      & $\dl 1.6(10)^n $ &               &                \\
  &              & \multicolumn{2}{c}{$-1.55(11)^o$}    &              &                      & $\dl 1.5(10)^o $ &               &                \\
34& $1.544(10) $ & $    -2.414(11) $ & $   -2.406(11) $ & $3.902(12) $ & $3.903(12) $         & $\dl 2.449(10) $ & $-8.703(11) $ & $-1.243(12) $  \\
  &              & \multicolumn{2}{c}{$-2.6(11)^n \,\;$}&              &                      & $\dl 2.6(10)^n $ &               &                \\
  &              & \multicolumn{2}{c}{$-3.03(11)^o$}    &              &                      & $\dl 3.3(10)^o $ &               &                \\
35& $1.868(10) $ & $    -2.977(11) $ & $   -2.967(11) $ & $4.782(12) $ & $4.782(12) $         & $\dl 3.024(10) $ & $-1.066(12) $ & $-1.523(12) $  \\
40& $4.490(10) $ & $    -7.799(11) $ & $   -7.772(11) $ & $1.219(13) $ & $1.220(13) $         & $\dl 7.968(10) $ & $-2.720(12) $ & $-3.884(12) $  \\
41& $5.281(10) $ & $    -9.316(11) $ & $   -9.283(11) $ & $1.450(13) $ & $1.450(13) $         & $\dl 9.527(10) $ & $-3.233(12) $ & $-4.618(12) $  \\
  &              & \multicolumn{2}{c}{$-1.05(12)^n$}    &              &                      &                  &               &                \\
  &              & \multicolumn{2}{c}{$-1.31(12)^o$}    &              &                      &                  &               &                \\
45& $9.735(10) $ & $    -1.818(12) $ & $   -1.812(12) $ & $2.784(13) $ & $2.784(13) $         & $\dl 1.866(11) $ & $-6.208(12) $ & $-8.866(12) $  \\
50& $1.946(11) $ & $    -3.869(12) $ & $   -3.855(12) $ & $5.824(13) $ & $5.826(13) $         & $\dl 3.983(11) $ & $-1.299(13) $ & $-1.855(13) $  \\
55& $3.645(11) $ & $    -7.647(12) $ & $   -7.620(12) $ & $1.136(14) $ & $1.136(14) $         & $\dl 7.894(11) $ & $-2.533(13) $ & $-3.617(13) $  \\
\hline \hline
\multicolumn{9}{l}{$(\ast)$     $\alpha_0(24S_{1/2})-\alpha_0(23S_{1/2})=0.37(9)$ Ref.~\cite{Fabre_Haroche_1978} (theo.); 
$(\ast\ast)$ $\alpha_0(24S_{1/2})-\alpha_0(23S_{1/2})=0.40(9)$ Ref.~\cite{Fabre_Haroche_1978} (exp.);}   \\
\multicolumn{9}{l}{$^a$Ref.~\cite{Maroulis_2004}; $^b$Ref.~\cite{Zhang_2007}; $^c$Ref.~\cite{Kamenski_2006}; $^d$Ref.~\cite{Ekstrom_1995}; $^e$Ref.~\cite{Zhu_2004}; $^f$Ref.~\cite{Molof_1974}; $^g$Ref.~\cite{Hannaford_1979}; $^h$Ref.~\cite{Harvey_1975} (theo.); $^i$Ref.~\cite{Harvey_1975} (exp.); } \\
\multicolumn{9}{l}{$^j$Ref.~\cite{Fabre_1975} (theo.); $^k$Ref.~\cite{Fabre_1975} (exp.); $^l$Ref.~\cite{Gallagher_1977} (theo.); $^m$Ref.~\cite{Gallagher_1977} (exp.); $^n$Ref.~\cite{Fabre_Haroche_1978} (theo.); $^o$Ref.~\cite{Fabre_Haroche_1978} (exp.)} 
\end{tabular}
\end{table*}

%########################################################################################

\begin{table*}[t] 
\caption {\label{tab:coeff} Determined expansion coefficients of Eq.~(\ref{eqn:coeff}) for the scalar and tensor polarizabilities of low- and high-lying states of~Na. The notation $(x)$ denotes $\times 10^x$.}
\centering
\begin{tabular}{cllll}
\hline\hline
$\alpha_{0,2}^{(n,l,j)}$  &  \multicolumn{1}{c}{$A_{0,2}^{(l,j)}$}   &   \multicolumn{1}{c}{$B_{0,2}^{(l,j)}$}  &  \multicolumn{1}{c}{$C_{0,2}^{(l,j)}$}  & \multicolumn{1}{c}{$\delta_{0,2}^{(l,j)}$} \\
\hline
$\alpha_{0} (nS_{1/2})$  & $\dl 1.21500(-1)$  & $\dl 7.26440(1)$  & $\dl 4.30899(1)$  & $\dl 1.43461$     \\
$\alpha_{0} (nP_{1/2})$  &    $-5.77239$      &    $-4.09439$     & $\dl 3.53175$     & $\dl 4.87711(-1)$ \\
$\alpha_{0} (nP_{3/2})$  &    $-5.75251$      &    $-2.45217$     &    $-9.71749(-1)$ & $\dl 7.22962(-1)$ \\
$\alpha_{2} (nP_{3/2})$  & $\dl 6.11501(-1)$  & $\dl 7.99240$     & $\dl 2.26122(1)$  & $\dl 2.40946$     \\
$\alpha_{0} (nD_{3/2})$  & $\dl 7.47735(1)$   & $\dl 7.89515$     & $\dl 1.92152(1)$  & $\dl 1.12710$     \\
$\alpha_{2} (nD_{3/2})$  &    $-1.66751(1)$   & $\dl 7.56835(-2)$ &    $-7.35366$     & $\dl 1.07230(-2)$ \\
$\alpha_{0} (nD_{5/2})$  & $\dl 7.47878(1)$   &    $-6.94563$     & $\dl 1.30233(1)$  &    $-9.93368(-1)$ \\
$\alpha_{2} (nD_{5/2})$  &    $-2.38141(1)$   & $\dl 9.20926(-2)$ &    $-7.36304$     & $\dl 1.30110(-2)$ \\
\hline\hline
\end{tabular}
\end{table*}

\begin{figure*}[h]  
\includegraphics[width=8.1cm]{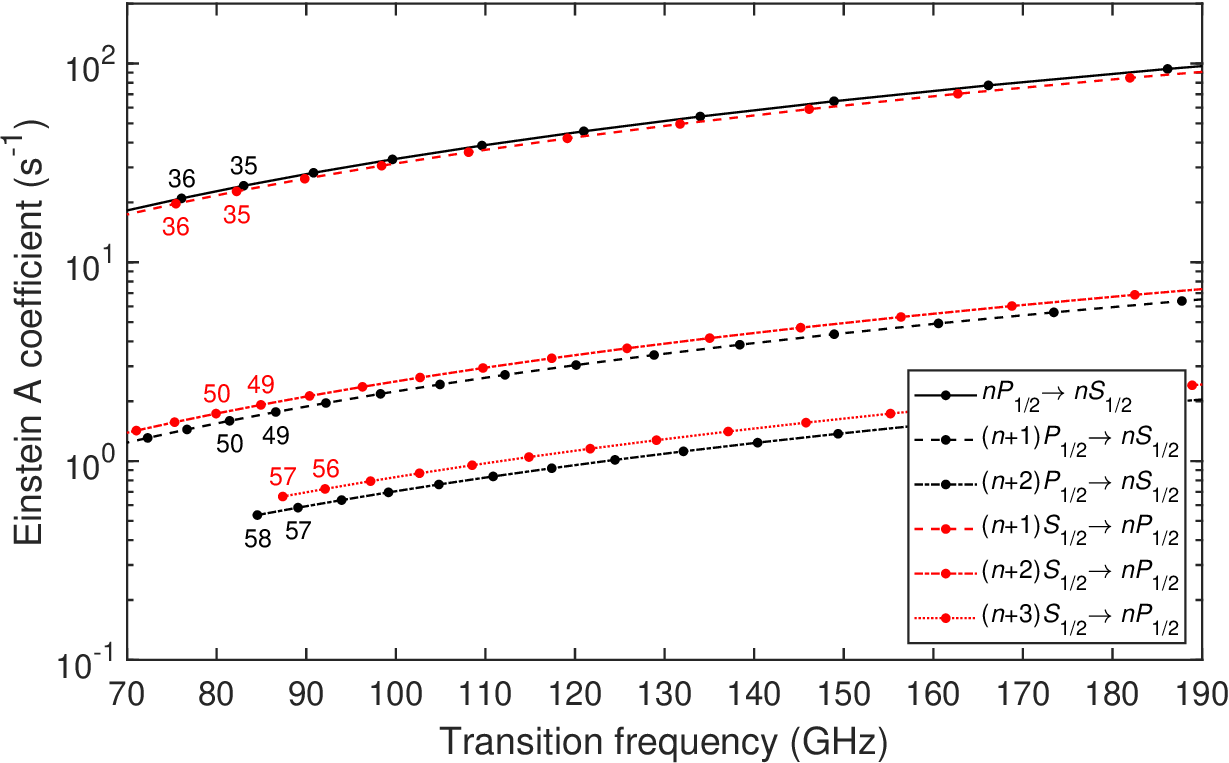} \hspace{0.5cm}
\includegraphics[width=8.1cm]{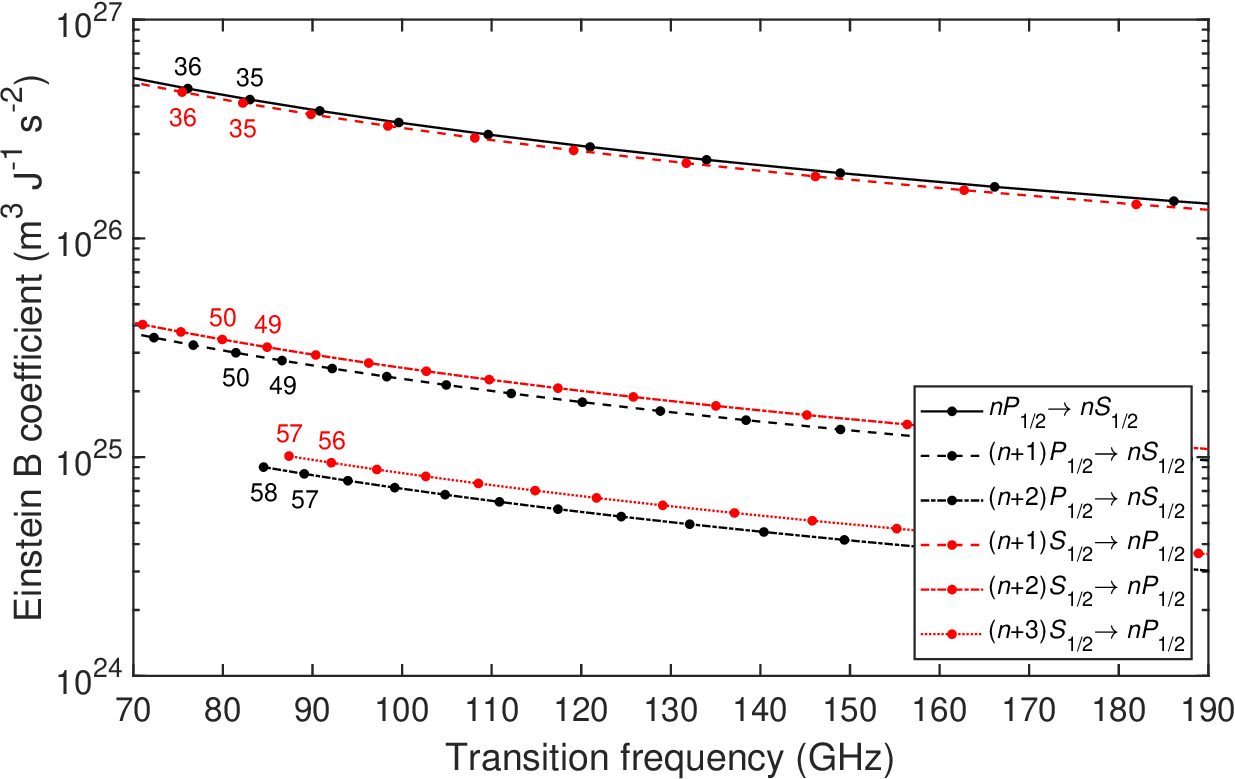} \\
\vspace{0.3cm}
\includegraphics[width=8.1cm]{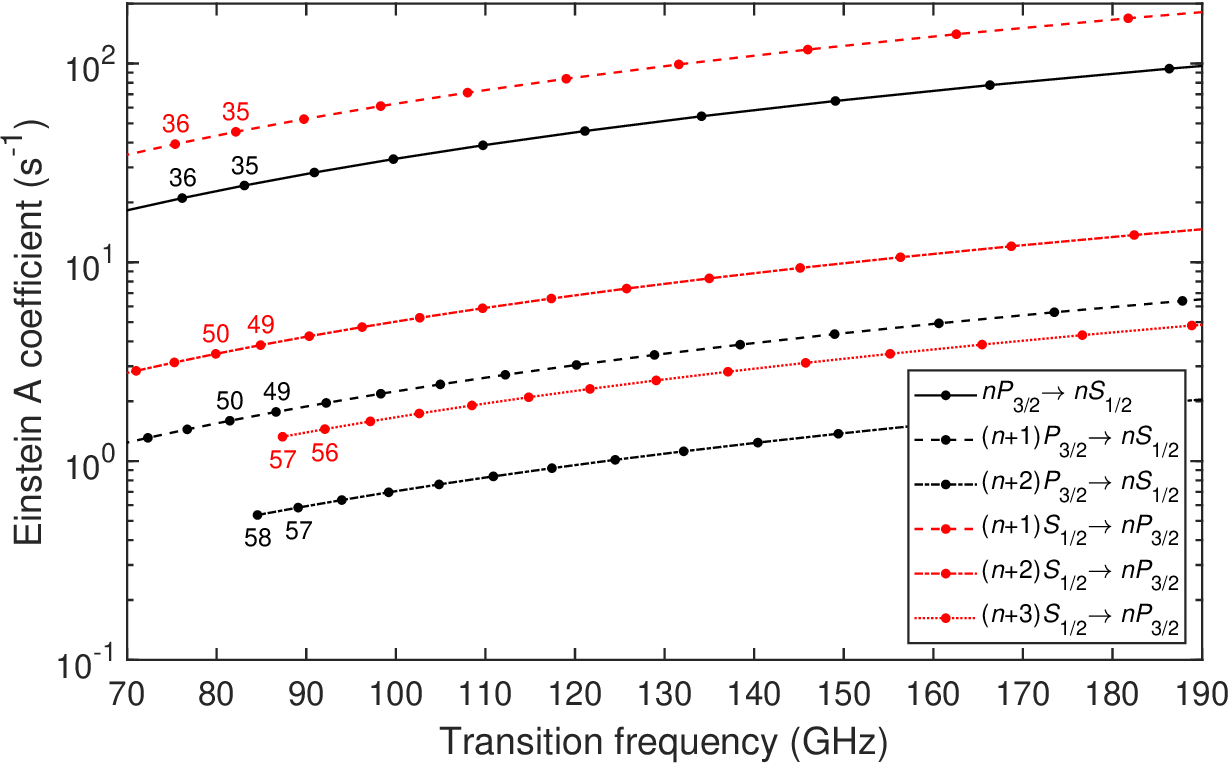} \hspace{0.5cm}
\includegraphics[width=8.1cm]{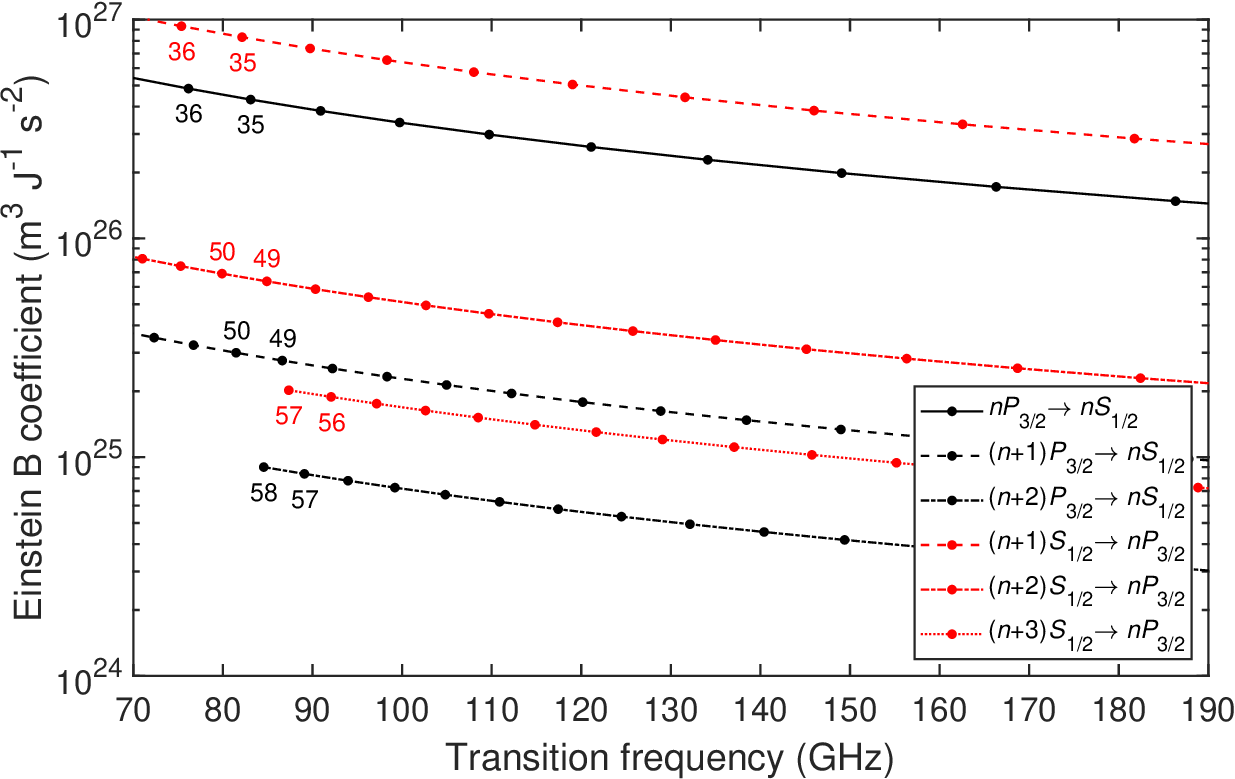} 
\caption{\label{fig:ABcoeff_SP} 
The largest Einstein $A$ (left panels) and $B$ (right panels) coefficients for the $nP_{1/2}\leftrightarrow n^\prime S_{1/2}$ (upper panels) and $nP_{3/2}\leftrightarrow n^\prime S_{1/2}$ (lower panels) microwave transitions among highly excited states in sodium. Each point correlates to a~specific principal quantum number, provided for some of the transitions noted in the legends.}
\end{figure*}

\begin{figure*}[t]
\includegraphics[width=8.1cm]{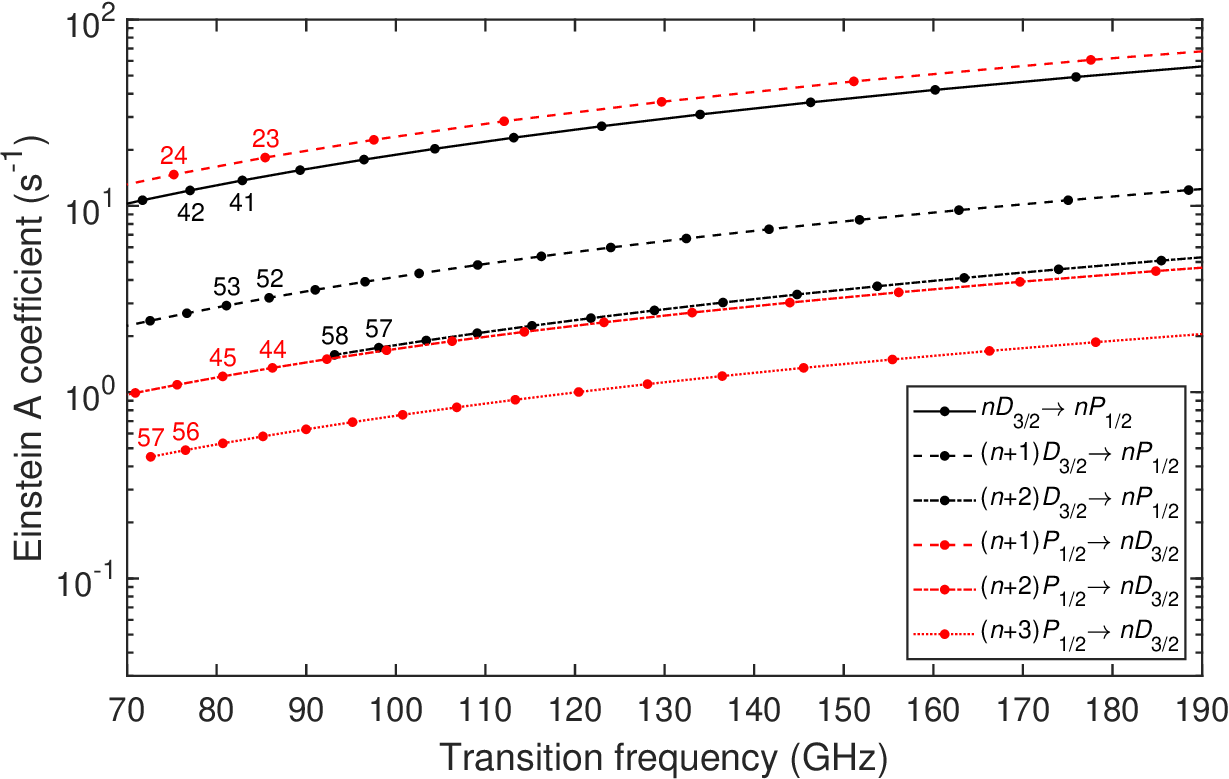} \hspace{0.5cm}
\includegraphics[width=8.1cm]{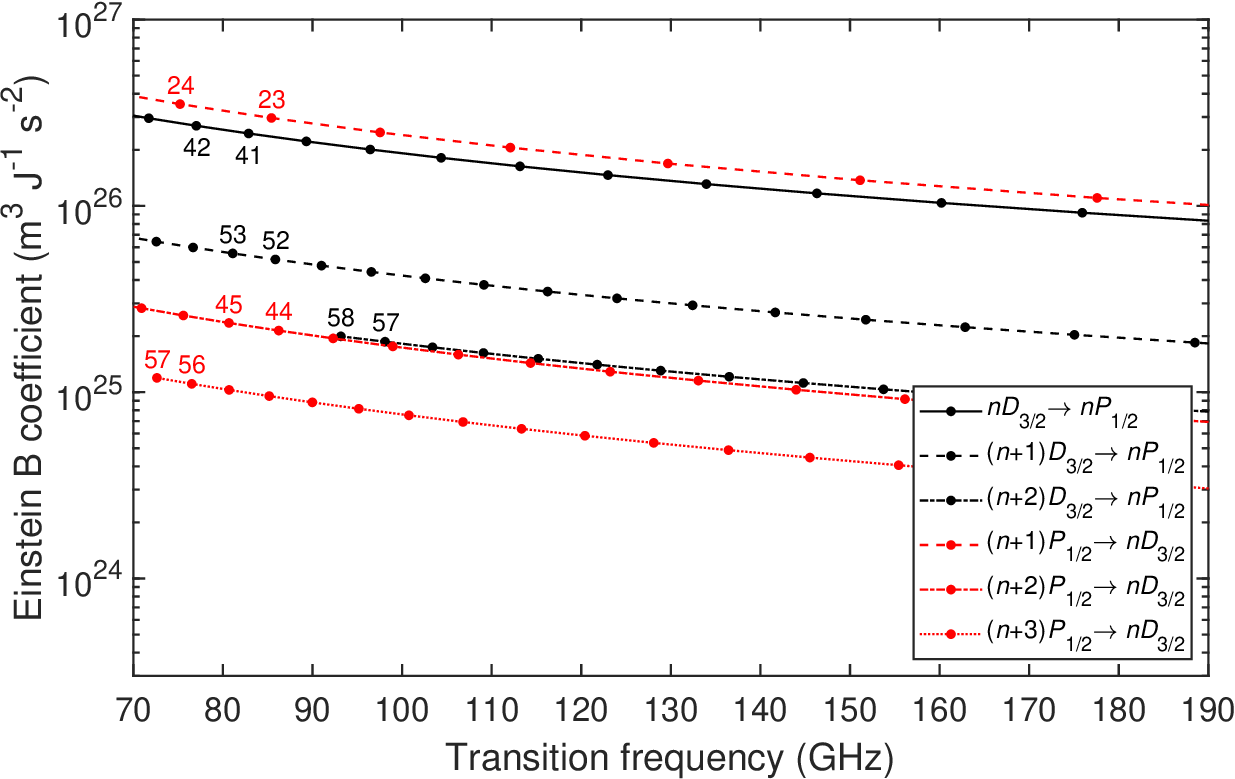} \\
\vspace{0.3cm}
\includegraphics[width=8.1cm]{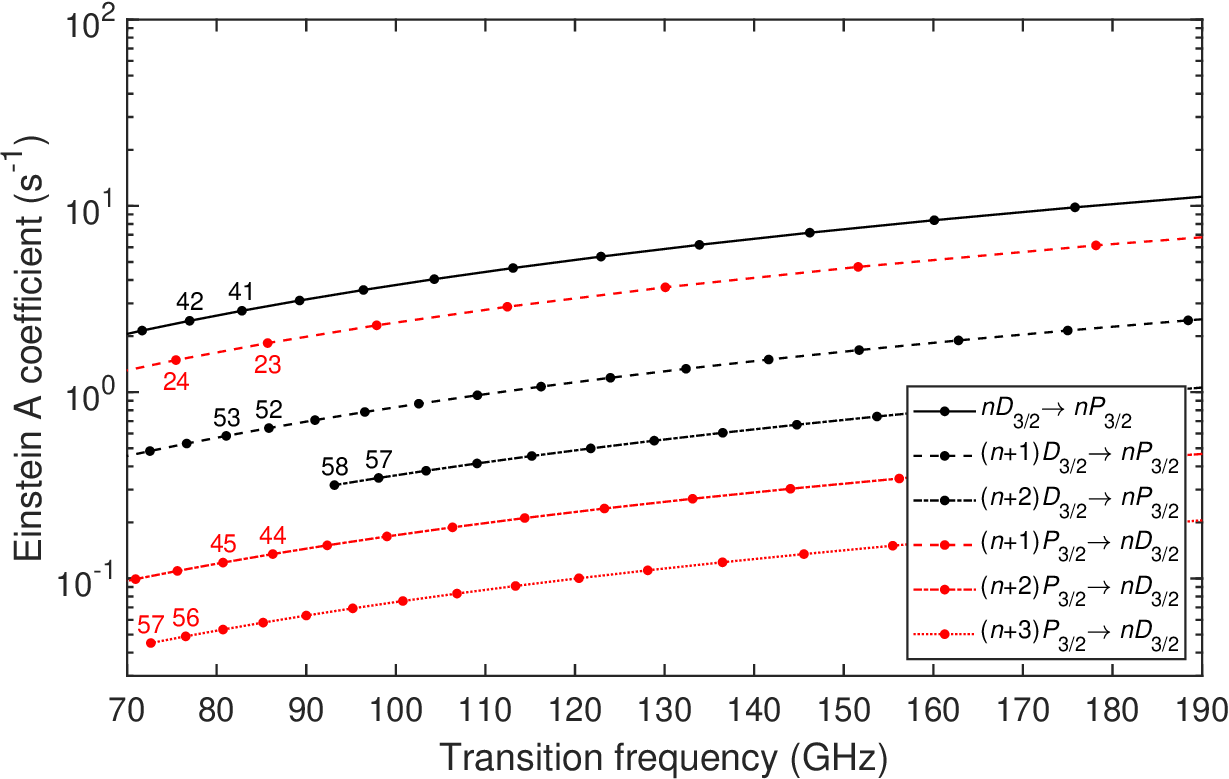} \hspace{0.5cm}
\includegraphics[width=8.1cm]{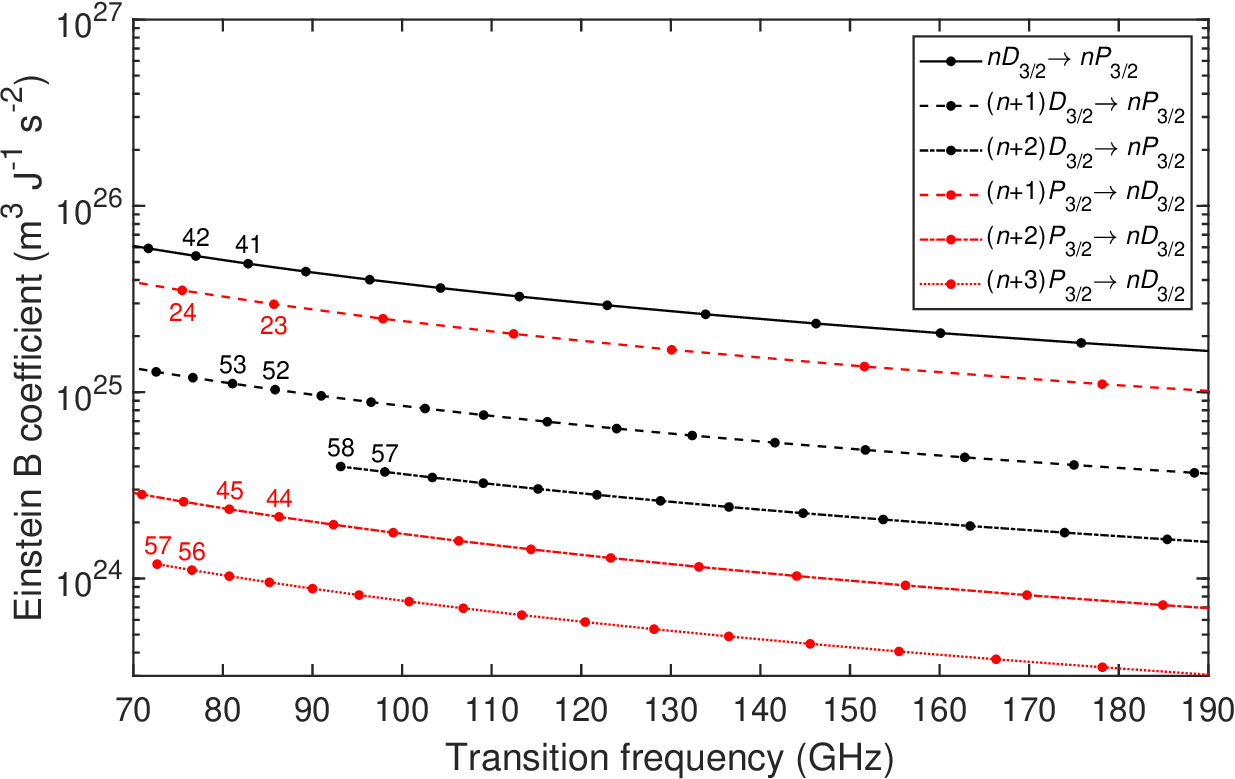} \\
\vspace{0.3cm}
\includegraphics[width=8.1cm]{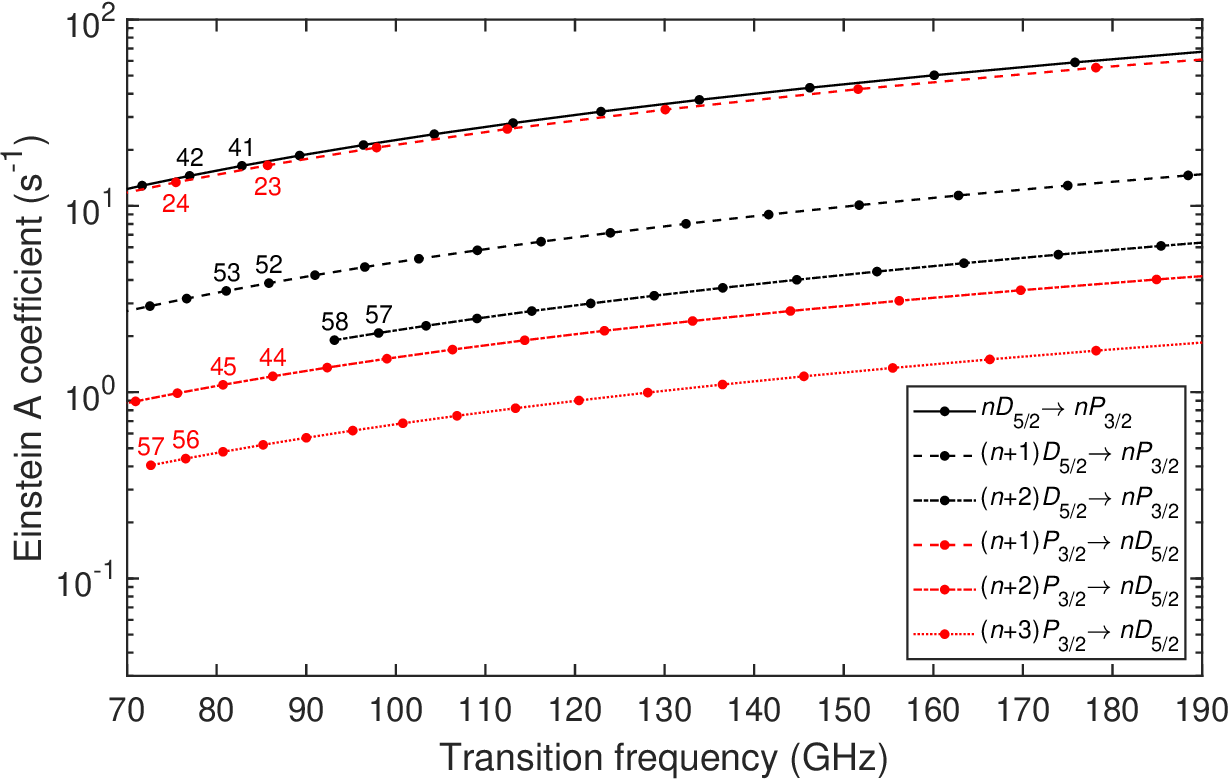} \hspace{0.5cm} 
\includegraphics[width=8.1cm]{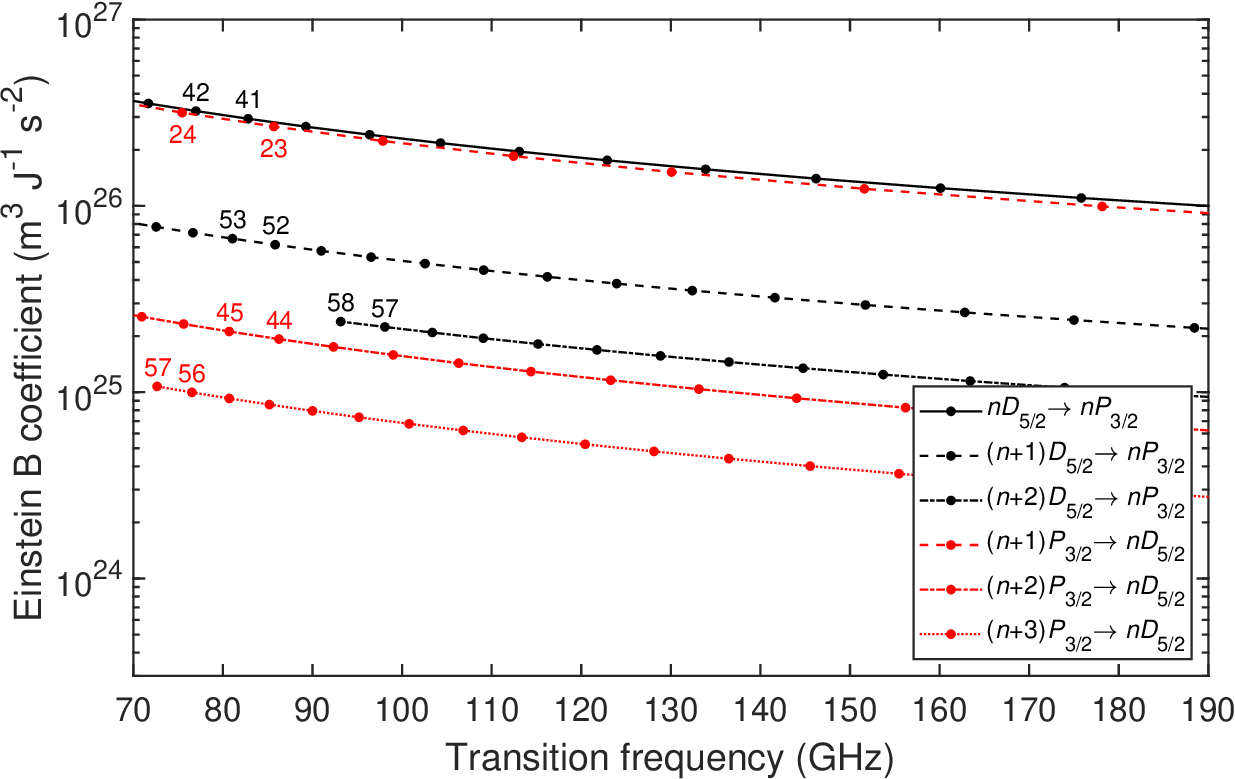}
\caption{\label{fig:ABcoeff_PD} 
The largest Einstein $A$ (left panels) and $B$ (right panels) coefficients for the $nD_{3/2}\leftrightarrow n^\prime P_{1/2}$ (upper panels), $nD_{3/2}\leftrightarrow n^\prime P_{3/2}$ (middle panels), and $nD_{5/2}\leftrightarrow n^\prime P_{3/2}$ (lower panels) microwave transitions among highly excited states in sodium. Each point correlates to a~specific principal quantum number, provided for some of the transitions noted in the legends.}
\vspace{0.5cm}   %%%
\end{figure*}

%%%%%%%%%%%%%%%%%%%%%%%%%%%%%%%%%%%%%%%%%%%%%%%%%%%%%%%%%%%%%%%%%%%%%%%%%%%%%
\section{Results and discussion} 

\subsection{Rydberg energy levels}
We calculate the energy levels for the $nS_{1/2}$, $nP_{1/2}$, $nP_{3/2}$, $nD_{3/2}$, $nD_{5/2}$, $nF_{5/2}$, and $nF_{7/2}$ series for $n\le 60$. Fig.~\ref{fig:band} presents energy level differences between two states in the 0--300~GHz range as a~function of the principal quantum number. The green area marks the 90--150~GHz microwave band. It is easy to determine between which two adjacent Rydberg states there is an atomic electron transition in a~specific frequency range; for example, the transition frequency of $nP_{1/2}\rightarrow nS_{1/2}$ is about 150~GHz for $n=29$ and 90~GHz for $n=34$. Fig.~\ref{fig:band} reveals that, in general, the one-photon transition for $|n-n'|\leq 2$ within the microwave band occurs when the principal quantum number is less than 60.

\subsection{Static polarizabilities of Rydberg states} 

The scalar and tensor polarizabilities for excited states of Na, an excellent test of the quality of the eigenfunctions and eigenvalues, are presented in Table~\ref{tab_polar}. The results are in good accord with theoretical~\cite{Gallagher_1977, Kamenski_2006,Zhu_2004, Fabre_Haroche_1978, Maroulis_2004, Zhang_2007, Harvey_1975, Fabre_1975} as well as experimental~\cite{Gallagher_1977, Fabre_Haroche_1978, Ekstrom_1995, Molof_1974, Hannaford_1979, Harvey_1975, Fabre_1975} findings. Although our basis set was optimized for excited energy levels, the calculated values for the lower lying levels are in good agreement with the observed values. Several different techniques have been employed to determine $\alpha_0$ and $\alpha_2$ for the ground state and the first low-lying $S$, $P$, and $D$ states; for example, Maroulis~\cite{Maroulis_2004} used the coupled cluster method with singles, doubles, and perturbative triples (CCSD(T)), whereas Zhu~\textit{et al.}~\cite{Zhu_2004} applied relativistic many-body perturbation theory. Kamenski and Ovsiannikov~\cite{Kamenski_2006} performed numerical computations with the Fues’ model potential. In turn, Zhang and Mitroy~\cite{Zhang_2007} combined the non-relativistic configuration interaction technique with the semiempirical core potential. The reference measurement of the ground-state polarizability of Na was done with an atom interferometer by Pritchard and coworkers~\cite{Ekstrom_1995}. Our value of $\alpha_0(S_{1/2})$ is in agreement to better than $\pm$3\% with the experimental finding and better than $\pm$1\% with the results of Maroulis~\cite{Maroulis_2004}, and of Kamenski and Ovsiannikov~\cite{Kamenski_2006}. It is interesting to point out that $\alpha_2(P_{3/2})$ for $n=4$, determined by the two latter authors, differs significantly from the result presented here by more than a~factor of two. Moreover, all the values of Kamenski and Ovsiannikov~\cite{Kamenski_2006} for $\alpha_2(P_{3/2})$ appear to be underestimated. A~less accurate experimental result for $\alpha_0(S_{1/2})$ comes from the work of Molof {\it et al.}~\cite{Molof_1974} using the $E$-$H$-gradient balance technique. In this case, the deviation is less than $\pm$5\%.

The typical experimental determination of polarizability is with Stark shift measurements of resonance transitions. However, the number of measurements for highly excited Rydberg states is relatively scarce. We contrasted our results with those of Fabre and Haroche~\cite{Fabre_1975} for $n= $ 10--12, Gallagher {\it et al.}~\cite{Gallagher_1977} for $n= $ 15--19, and Fabre {\it et al.}~\cite{Fabre_Haroche_1978} for $n= $ 23--41. When a weak electric field is applied, the tensor polarizability of an atom with one valence electron can be effectively obtained from the difference in polarizabilities between two proper magnetic sublevels. Assuming that the fine-structure splitting is small in comparison to the energy distance for $l= l^\prime \pm 1$, Eqs.~(\ref{eqn:scalar}) and (\ref{eqn:tensor}) simplify, which consequently leads to $\alpha_0(P_{1/2})$ = $\alpha_0(P_{3/2})$, $\alpha_0(D_{3/2})$ = $\alpha_0(D_{5/2})$, and $\alpha_2(D_{3/2})$ = $\frac{7}{10}\alpha_2(D_{5/2})$. Note that $\alpha_2(P_{1/2})$ is always equal to zero. Our results show that this is quite a reasonable approximation for the sodium atom, especially for the highly excited states. A~smaller, but still significant, difference between $\alpha_2(F_{5/2})$ and $\alpha_2(F_{7/2})$ also exists. This is because the spin--orbit splitting decreases with increasing $l$. Based on the above mentioned approximation, $\alpha_2(F_{5/2})$ will be equal to $\frac{6}{7}\alpha_2(F_{7/2})$. One can see in Table~\ref{tab_polar}, the striking agreement between our scalar and tensor polarizabilities for $n=32$ with the experimental values of Fabre {\it et al.}~\cite{Fabre_Haroche_1978}. The agreement is less impressive for higher excited states ($n=34$ and $n=41$). The polarizabilities were also determined theoretically, based on the Coulomb approximation for those states in~Ref.~\cite{Fabre_Haroche_1978}.

The Rydberg polarizabilities scale as $n^7$~\cite{Kamenski_2014}. To cover the entire space of bound states with the values of $\alpha_0$ and $\alpha_2$ for the sodium atom, we may expand the polarizability in terms of a~power series of the effective principal quantum number $n^{\ast}=n-\delta^{(l,j)}$, where $\delta^{(l,j)}$ is the quantum defect, as follows
\begin{eqnarray}
 \!\!\!\!\!\!\!\!\!   \alpha_{0,2}^{(n,l,j)} &=& A_{0,2}^{(l,j)} \left( n-\delta_{0,2}^{(l,j)} \right)^{\!7}  \nonumber \\ 
 &&    \times \left(1 + \frac{B_{0,2}^{(l,j)}}{ n-\delta_{0,2}^{(l,j)} }  +  \frac{C_{0,2}^{(l,j)}}{\left ( n-\delta_{0,2}^{(l,j)} \right )^2} + ...  \right)\!.  \label{eqn:coeff}
\end{eqnarray}
The fitted coefficients $A_{0,2}^{(l,j)}$, $B_{0,2}^{(l,j)}$, $C_{0,2}^{(l,j)}$, and $\delta_{0,2}^{(l,j)}$ of calculated  $\alpha_0$ and $\alpha_2$ with the above expression, are presented in Table~\ref{tab:coeff} for the $S_{1/2}$, $P_{1/2,3/2}$, and $D_{3/2,5/2}$ states. The goodness of the fits is excellent. The three-term expansion~(\ref{eqn:coeff}) with the coefficients, Table~\ref{tab:coeff}, reproduces the scalar and tensor polarizabilities to four significant figures, as shown in Table~\ref{tab_polar}.

%%%%%%%%%%%%%%%%%%%%%%%%%%%%%%%%%%%%%%%%%%%%%%%%%%%%%%%%%%%%%%%%%%%%%%%%%%%%%
\subsection{Einstein \textit{A} \& \textit{B} coefficients}

We determined the Einstein coefficients of the $nP_{1/2,3/2}\leftrightarrow n^\prime S_{1/2}$ and $nD_{3/2,5/2}\leftrightarrow n^\prime P_{1/2,3/2}$ microwave emissions in the sodium Rydberg series. The upper left panel of Fig.~\ref{fig:ABcoeff_SP} presents the results of the $A$~coefficients for the $nP_{1/2}\rightarrow nS_{1/2}$, $(n+1)P_{1/2}\rightarrow nS_{1/2}$, $(n+2)P_{1/2}\rightarrow nS_{1/2}$, $(n+1)S_{1/2}\rightarrow nP_{1/2}$, $(n+2)S_{1/2}\rightarrow nP_{1/2}$, and $(n+3)S_{1/2}\rightarrow nP_{1/2}$ transitions in the 70--190~GHz frequency range. The two most dominant $A$~coefficients are for the $nP_{1/2}\rightarrow nS_{1/2}$ and $(n+1)S_{1/2}\rightarrow nP_{1/2}$ one-photon decay, with $n=34$ at about 90~GHz and $n=29$ at about 150~GHz. The lower left panel of Fig.~\ref{fig:ABcoeff_SP} is for transitions emanating from and to $P_{3/2}$. It is worth noting that the $A$ coefficient is about two times larger for $nS_{1/2}\rightarrow n^\prime P_{3/2}$ than for $nS_{1/2}\rightarrow n^\prime P_{1/2}$. The ratio $\frac{A(nS_{1/2}\rightarrow n^\prime P_{3/2})}{A(nS_{1/2}\rightarrow n^\prime P_{1/2})}=2$, but only when one assumes $E_{n^\prime P_{1/2}}=E_{n^\prime P_{3/2}}$, i.e., when spin--orbit coupling is neglected.

\begin{table}[t]
\caption {\label{tab_A} Einstein $A$ coefficients ($10^5$ s$^{-1}$) for spontaneous emission to the Na($3P_{3/2}$) state.}
\begin{tabular}{cllll} 
\hline \hline
& \multicolumn{2}{c}{$nS_{1/2}\rightarrow 3P_{3/2}$}
& \multicolumn{2}{c}{$nD_{3/2}\rightarrow 3P_{3/2}$}\\
$n$ & \multicolumn{1}{c}{Other works} &  \multicolumn{1}{c}{This work} &  \multicolumn{1}{c}{Other works} &  \multicolumn{1}{c}{This work} \\
\hline 
3  &                    &             &  85.76$^a$       &  85.59      \\
   &                    &             &  84.95$^b$       &             \\
   &                    &             &  85.7$^c$        &             \\
4  & 178.0$^a$          & 175.0       &  20.23$^a$       &  20.50      \\
   & 176$^c$            &             &  20.15$^b$       &             \\
   &                    &             &  20.2$^c$        &             \\
5  & \dd 50.75$^a$      & \dd 49.50   & \dd 8.147$^a$    & \dd 8.298   \\
   & \dd 49.8$^c$       &             & \dd 8.109$^b$    &             \\
   &                    &             & \dd 8.15$^c$     &             \\
6  & \dd 23.15$^a$      & \dd 22.51   & \dd 4.159$^a$    & \dd 4.244   \\
   & \dd 22.7$^c$       &             & \dd 4.129$^b$    &             \\
   &                    &             & \dd 4.14$^c$     &             \\
7  & \dd 12.63$^a$      & \dd 12.24   & \dd 2.429$^a$    & \dd 2.484   \\
   & \dd 12.3$^c$       &             & \dd 2.44$^c$     &             \\
8  & \dd\dd 7.662$^a$   & \dd 7.414   & \dd 1.550$^a$    & \dd 1.587   \\
   & \dd\dd 7.50$^c$    &             & \dd 1.95$^c$     &             \\
9  & \dd\dd 5.000$^a$   & \dd 4.834   & \dd 1.141$^a$    & \dd 1.079   \\
   & \dd\dd 5.61$^c$    &             &                  &             \\
10 & \dd\dd 3.444$^a$   & \dd 3.328   & \dd 0.7498$^a$   & \dd 0.7688  \\
   & \dd\dd 6.50$^c$    &             &                  &             \\
11 & \dd\dd 2.473$^a$   & \dd 2.390   & \dd 0.5535$^a$   & \dd 0.5678  \\
12 & \dd\dd 1.836$^a$   & \dd 1.774   & \dd 0.4200$^a$   & \dd 0.4317  \\
13 & \dd\dd 1.401$^a$   & \dd 1.353   & \dd 0.3262$^a$   & \dd 0.3362  \\
14 & \dd\dd 1.094$^a$   & \dd 1.055   & \dd 0.2587$^a$   & \dd 0.2670  \\
15 & \dd\dd 0.8700$^a$  & \dd 0.8391  & \dd 0.2088$^a$   & \dd 0.2157  \\
16 & \dd\dd 0.7032$^a$  & \dd 0.6782  & \dd 0.1710$^a$   & \dd 0.1768  \\
17 & \dd\dd 0.5764$^a$  & \dd 0.5560  & \dd 0.1418$^a$   & \dd 0.1467  \\
18 & \dd\dd 0.4783$^a$  & \dd 0.4615  & \dd 0.1189$^a$   & \dd 0.1231  \\
19 & \dd\dd 0.4012$^a$  & \dd 0.3872  & \dd 0.1008$^a$   & \dd 0.1044  \\
20 & \dd\dd 0.3398$^a$  & \dd 0.3281  & \dd 0.08616$^a$  & \dd 0.08925 \\
21 & \dd\dd 0.2904$^a$  & \dd 0.2804  & \dd 0.07426$^a$  & \dd 0.07692 \\
22 & \dd\dd 0.2500$^a$  & \dd 0.2416  & \dd 0.06446$^a$  & \dd 0.06677 \\
23 & \dd\dd 0.2169$^a$  & \dd 0.2096  & \dd 0.05630$^a$  & \dd 0.05833 \\
24 & \dd\dd 0.1894$^a$  & \dd 0.1830  & \dd 0.04946$^a$  & \dd 0.05126 \\
25 & \dd\dd 0.1663$^a$  & \dd 0.1607  & \dd 0.04369$^a$  & \dd 0.04529 \\
26 & \dd\dd 0.1469$^a$  & \dd 0.1419  & \dd 0.03879$^a$  & \dd 0.04021 \\
27 & \dd\dd 0.1304$^a$  & \dd 0.1259  & \dd 0.03459$^a$  & \dd 0.03587 \\
28 & \dd\dd 0.1162$^a$  & \dd 0.1123  & \dd 0.03098$^a$  & \dd 0.03213 \\
29 & \dd\dd 0.1041$^a$  & \dd 0.1005  & \dd 0.02786$^a$  & \dd 0.02889 \\
30 & \dd\dd 0.09356$^a$ & \dd 0.09035 & \dd 0.02514$^a$  & \dd 0.02608 \\
35 & \dd\dd 0.05775$^a$ & \dd 0.05574 & \dd 0.01578$^a$  & \dd 0.01637 \\
40 & \dd\dd 0.03811$^a$ & \dd 0.03678 & \dd 0.01055$^a$  & \dd 0.01095 \\
45 & \dd\dd 0.02646$^a$ & \dd 0.02553 & \dd 0.007402$^a$ & \dd 0.007678 \\
50 & \dd\dd 0.01911$^a$ & \dd 0.01844 & \dd 0.005392$^a$ & \dd 0.005592 \\
55 &                    & \dd 0.01375 &                  & \dd 0.004198 \\
60 & \dd\dd             & \dd 0.01052 &                  & \dd 0.003223 \\
\hline \hline
\multicolumn{5}{l}{$^a$Ref.~\cite{Miculis_2005};  $^b$Ref.~\cite{Laughlin_1992} (taken from Ref.~\cite{Miculis_2005}); $^c$Ref.~\cite{Kelleher}} 
\end{tabular}
\end{table}

\begin{table}[t]
\caption{\label{lifetimefig} Lifetimes (in $\mu$s) of sodium Rydberg states, compared with low-temperature measurements.} 
\centering
\begin{tabular}{c@{\qquad}l@{\qquad}cc}
\hline\hline
$n$ & \multicolumn{1}{l}{$nS_{1/2}$} & \multicolumn{1}{c}{$nD_{3/2}$} & \multicolumn{1}{c}{$nD_{5/2}$} \\
\hline
17 &  \dd 5.203    & 4.508 &  4.512                \\
   &               & \multicolumn{2}{c}{4.46$^a$}  \\
18 &  \dd 6.263    & 5.351 &  5.356                \\
   &               & \multicolumn{2}{c}{5.75$^a$}  \\
19 &  \dd 7.459    & 6.294 &  6.299                \\
   &  \dd 7.42$^a$ & \multicolumn{2}{c}{6.90$^a$}  \\
20 &  \dd 8.800    & 7.340 &  7.346                \\
   &  \dd 8.9$^a$  & \multicolumn{2}{c}{7.7$^a$}   \\
21 &  10.29        & 8.496 &  8.503                \\
   &  11.3$^a$     & \multicolumn{2}{c}{8.6$^a$}   \\
22 &  11.94        & 9.768 &  9.776                \\
   &  12.2$^a$     & \multicolumn{2}{c}{10.2$^a$}  \\
23 &  13.76        & 11.16 & 11.17                 \\
   &  14.5$^a$     & \multicolumn{2}{c}{11.4$^a$}  \\
24 &  15.76        & 12.68 & 12.69                 \\
   &  16.6$^a$     & \multicolumn{2}{c}{13.9$^a$}  \\
25 &  17.94        & 14.33 & 14.34                 \\
   &  18.6$^a$     & \multicolumn{2}{c}{15.0$^a$}  \\
26 &  20.32        & 16.12 & 16.13                 \\
   &  21.2$^a$     & \multicolumn{2}{c}{16.9$^a$}  \\
27 &  22.89        & 18.05 & 18.06                 \\
   &  23.8$^a$     & \multicolumn{2}{c}{17.7$^a$}  \\
28 &  25.67        & 20.12 & 20.14                 \\
   &  24.9$^a$   \\  
\hline\hline
\multicolumn{4}{l}{$^a$Ref.~\cite{Spencer_1981}}
\end{tabular}
\end{table}

\begin{figure}[b]
\includegraphics[width=8.0cm]{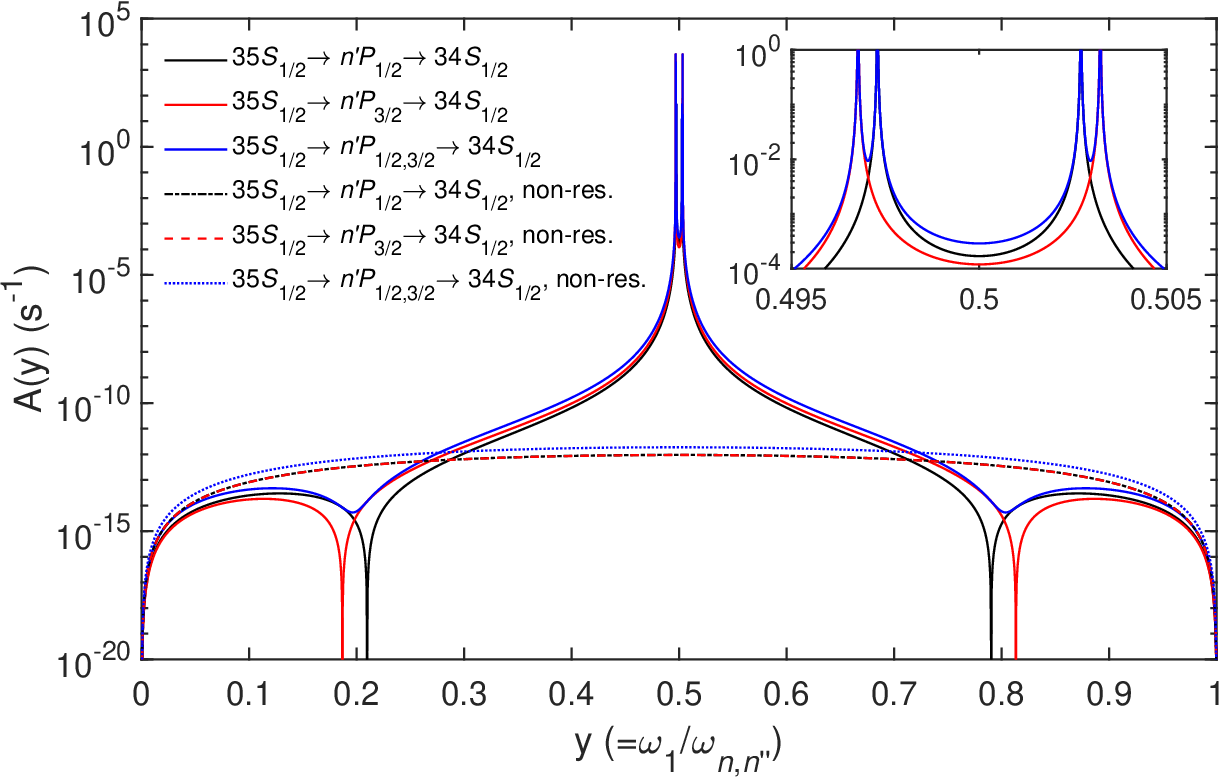} \\
\vspace{0.3cm}
\includegraphics[width=8.0cm]{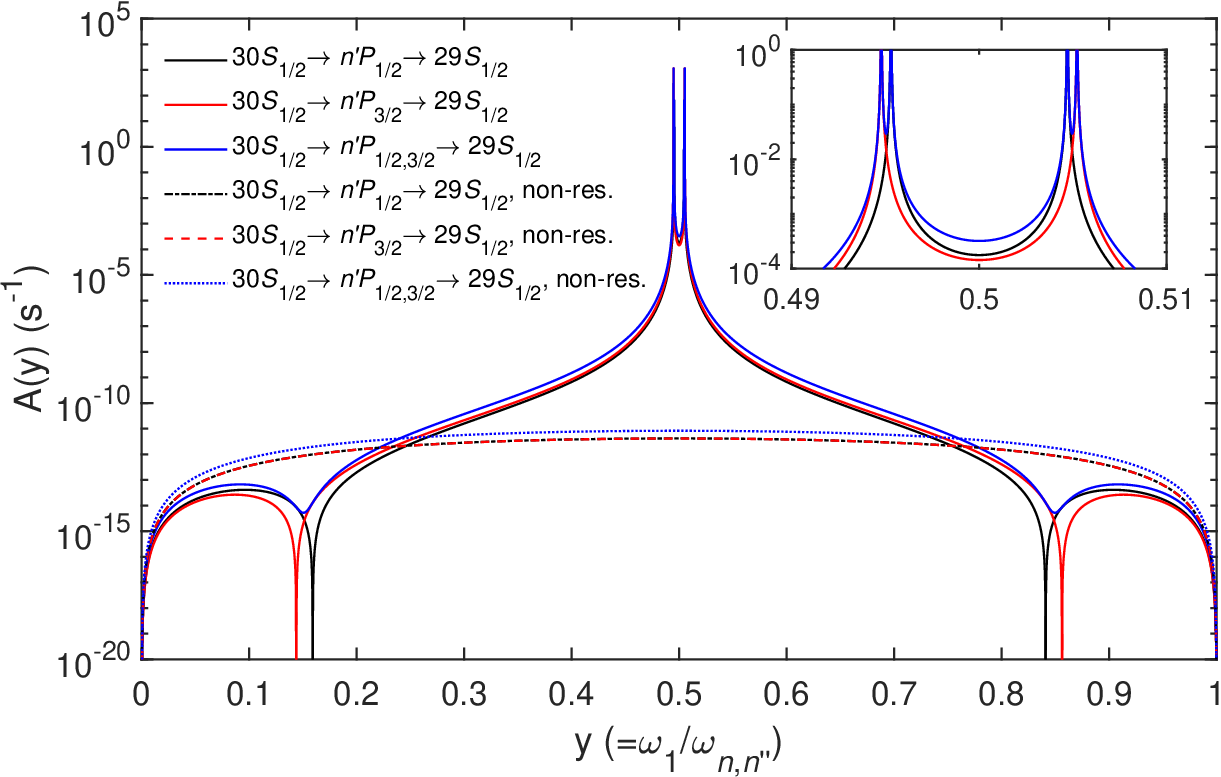}
\caption{\label{fig:Two_photon} 
Two-photon emission spectra for the $35S_{1/2}\rightarrow34S_{1/2}$ (upper panel) and $30S_{1/2}\rightarrow29S_{1/2}$ (lower panel) transitions. The corresponding frequencies $\omega_{n,n''}/2\pi$ are 181 and 295~GHz, respectively. The variable $y$ is the single-photon frequency $\omega_1$ with respect to the two-photon transition frequency $\omega_{n,n''} (=\omega_1+\omega_2)$. The black and red lines represent the transitions via $n^\prime P_{1/2}$ and $n^\prime P_{3/2}$, respectively, whereas the blue line is the sum of these two. The dotted-dashed, dashed, and dotted lines represent the non-resonant spectra.}
\end{figure}

Fig.~\ref{fig:ABcoeff_PD} shows the Einstein $A$ coefficients for the $nD_{3/2}\leftrightarrow n^\prime P_{1/2}$ (upper left panel), $nD_{3/2}\leftrightarrow n^\prime P_{3/2}$ (middle left panel), and $nD_{5/2}\leftrightarrow n^\prime P_{3/2}$ (lower left panel) microwave transitions, where $|n - n^\prime|\leq 3$. Figures~\ref{fig:ABcoeff_SP} and~\ref{fig:ABcoeff_PD} allow an easy reading between exactly which states the transition occurs at a~given frequency --- each point corresponds to a~specific $n$. The largest $A$~coefficient among those investigated here is for the $(n+1)S_{1/2}\rightarrow nP_{3/2}$ decay. An~analogous situation occurs for the $B$~coefficients, which are displayed in the right panels in Figs.~\ref{fig:ABcoeff_SP} and \ref{fig:ABcoeff_PD}. It should be noted that as the frequency increases, each of the curves for the $B$~coefficients tends to trend lower.

The $A$ coefficients for spontaneous emission from the low- and high-lying $nS_{1/2}$ and $nD_{3/2}$ energy levels, where $n\leq 60$, to the lowest $P_{3/2}$ state are presented in Table~\ref{tab_A}. They are compared with theoretical results of Miculis and Meyer~\cite{Miculis_2005} and the available NIST data~\cite{Kelleher}. The agreement is excellent. Note that the report of Kelleher and Podobedova from NIST~\cite{Kelleher} contains data for transitions from low-excited states only. We would like to point out one particular transition in the NIST report, i.e.,  the $10S_{1/2}\rightarrow 3P_{3/2}$ transition. It is almost double our result as well as the tranisiton of Miculis and Meyer~\cite{Miculis_2005}.

Additionally, we determined the radiative lifetimes of the sodium $nS_{1/2}$ and $nD_{3/2,5/2}$ states for $n=17$--28 and compared them directly with low-temperature measurements of Spencer {\it et al.}~\cite{Spencer_1981}. The lifetime $(\tau_k)$ against spontaneous emission of the upper state $(k)$ is related to the sum of such Einstein $A_{k,i}$ coefficients over all allowed transitions to lower states, $\tau^{-1}_k=\sum_i A_{k,i}$. Since the experiment was carried out in a~cooled environment, the effect of blackbody-induced transfer can be neglected in the theoretical considerations. Our results and the experimental values are presented in Table~\ref{lifetimefig}. One may notice the measurement data exhibit less smooth behavior than the computed values; however, the agreement between the results is good. The lifetime increases with increasing $n$, varying between 4 and 26 $\mu$s.

%%%%%%%%%%%%%%%%%%%%%%%%%%%%%%%%%%%%%%%%%%%%%%%%%%%%%%%%%%%%%%%%%%%%%%%%%%%%%
\subsection{Two-photon transitions}
Dyubko and coworkers~\cite{Dyubko_1995} measured the frequency of the two-photon transition Na($29S_{1/2} \rightarrow 30S_{1/2}$) and obtained a~value of $2\times 147542$~MHz. Our calculated value is $2\times 147528$~MHz, helping to validate the accuracy of our results. The fractional difference is less than 0.01\%.

We calculated the full two-photon emission spectra for transitions $35S_{1/2}$ to $34S_{1/2}$ and  $30S_{1/2}$ to $29S_{1/2}$ via $n^\prime P_{j^\prime}$. The choice was dictated by the fact that $E_{35S_{1/2}}-E_{34S_{1/2}}=181$~GHz ($\approx 2\times 90$~GHz) and $E_{30S_{1/2}}-E_{29S_{1/2}}=295$~GHz ($\approx 2 \times 150$~GHz); moreover, as we mentioned earlier, the highest values for the  one-photon emission probabilities are for the $nP_{j}\rightarrow nS_{1/2}$ and $(n+1)S_{1/2}\rightarrow nP_{j^\prime}$ transitions. The results are presented in Fig.~\ref{fig:Two_photon}. The black and red curves show the transitions through the $n'P_{1/2}$ and $n'P_{3/2}$ states, respectively. The sum of the two transitions, as the probability of two-photon decay, is shown in blue. The sharp and nearly Lorentzian peaks correspond to $n^\prime P_{j^\prime}$ resonances, shifted, due to the spin--orbit interaction, as is clearly visible in the insets of Fig.~\ref{fig:Two_photon}. Furthermore, one sees the black and red curves suddenly drop to zero at certain frequencies. This behavior is caused by a destructive interference, resulting in the full cancellation of $M_{j^\prime}^{(1)}$ and $M_{j^\prime}^{(2)}$ terms, see Eq.~(\ref{lab_Mj}), at specific values of~$y$. In experiments, due to inhomogeneity and broadening, $A(y)$ will never be zero, but in our calculations, there is no other width than the natural linewidth.

Interestingly, the probability of emission of two identical photons is non-zero. The transition rate for $(n+1)S_{1/2}\rightarrow nS_{1/2}$ at $y=0.5$ $(\omega_1=\omega_2=\omega_{n,n''}/2)$ is equal to $1.70 \times 10^{-4}$, $1.20 \times 10^{-4}$, $1.77 \times 10^{-4}$, and $1.44 \times 10^{-4}$~s$^{-1}$, respectively, for the pair of quantum numbers $(n,j^\prime)$: $(34,\frac{1}{2})$, $(34,\frac{3}{2})$, $(29,\frac{1}{2})$, and $(29,\frac{3}{2})$.

We also investigated the non-resonant contribution to the two-photon spectral distribution $A(y)$, carrying out the summation in Eq.~(\ref{lab_Mj}) over all states with energies $E_{n'}\geq E_{n}$ and $E_{n'}\leq E_{n''}$. One can identify, in Fig.~\ref{fig:Two_photon}, that it is small on the absolute scale compared with the peaks. However, the non-resonant terms noticeably increase the distant wings of the two-photon emission profiles (for about $y=0$ and $y=1$). After numerical integration, we obtained 
$A_t^{\rm non\textrm{-}res.} (35S_{1/2} \rightarrow n^\prime P_{1/2} \rightarrow 34S_{1/2}$ and $n^\prime\neq 34) = 2.42\times 10^{-13}$~s$^{-1}$, 
$A_t^{\rm non\textrm{-}res.} (35S_{1/2} \rightarrow n^\prime P_{3/2} \rightarrow 34S_{1/2}$ and $n^\prime\neq 34) = 2.39\times 10^{-13}$~s$^{-1}$,
$A_t^{\rm non\textrm{-}res.} (30S_{1/2} \rightarrow n^\prime P_{1/2} \rightarrow 29S_{1/2}$ and $n^\prime\neq 29) = 1.06\times 10^{-12}$~s$^{-1}$, and 
$A_t^{\rm non\textrm{-}res.} (30S_{1/2} \rightarrow n^\prime P_{3/2} \rightarrow 29S_{1/2}$ and $n^\prime\neq 29) = 1.05\times 10^{-12}$~s$^{-1}$.
Of course, within the non-relativistic treatment for the two-photon transition from $(n+1)S$ to $nS$ of hydrogen-like atoms, only non-resonant contributions to the total two-photon decay rate matter because the energy levels of the same $n$ but different $l$ are degenerate. There are no real intermediate states; consequently, resonant contributions do not occur~\cite{Chluba_2008}, in contrast to our case.

%%%%%%%%%%%%%%%%%%%%%%%%%%%%%%%%%%%%%%%%%%%%%%%%%%%%%%%%%%%%%%%%%%%%%%%%%%%%%
\section{Conclusions}
We have accurately calculated the highly excited $S$, $P$, $D$, and $F$ Rydberg states of the sodium atom within the Ritz variational method. The positive core--electron interaction was modeled by the parametric one-electron valence potential with spin-orbit coupling. We have determined the scalar and tensor polarizabilities for low and highly excited states and compared them with available experimental and theoretical results. After representing the polarizability as a~series in powers of the effective principal quantum number, we found the leading expansion coefficients, allowing for rapid evaluation of $\alpha_0$ and $\alpha_2$ for arbitrary $n$ and $l\le 2$. As recently shown, utilizing the enormous polarizabilities of Rydberg states with the control of long-range interaction enables one to image the dynamics of ions embedded in a cold cloud of atoms~\cite{Gross_2020}. We also calculated the largest Einstein coefficients for spontaneous and stimulated emission in the band of interest for photometric observations. From Fig.~\ref{fig:ABcoeff_SP} and \ref{fig:ABcoeff_PD}, one can immediately determine between which states the transitions take place for a~given frequency. In general, the most probable Na transitions occur from the $(n,l,j)$ state to the $(n,l-1,j-1)$ and $(n-1,l+1,j+1)$ states. We have also shown the emission of photons due to two-photon transitions from $30S_{1/2}$ to $29S_{1/2}$ and from $35S_{1/2}$ to $34S_{1/2}$.

The emphasis in our studies was on transitions in the~microwave frequency range between 90 and 150~GHz. They may be helpful in calculating the expected LPRS signal flux density on Earth's surface at various microwave frequencies from the resulting laser-excited source in the sodium layer as a~function of power and other properties of the two lasers and for precise relative radiometric and polarimetric calibration of ground-based microwave and millimeter-wave telescopes.

%%%%%%%%%%%%%%%%%%%%%%%%%%%%%%%%%%%%%%%%%%%%%%%%%%%%%%%%%%%%%%%%%%%%%%%%%%%%%
\section*{Acknowledgements}
This work was partially supported at ITAMP by an NSF grant to Harvard University.

%\bibliography{papers}


\end{document}